\documentclass[a4paper]{article}

\usepackage{a4wide}
\usepackage{epstopdf}
\usepackage{subfigure}

\usepackage{amsthm}
\usepackage{graphicx}

\usepackage{booktabs}
\usepackage{graphicx}
\usepackage{dcolumn}
\usepackage{color}
\usepackage{float}
\usepackage{natbib}
\usepackage{multirow}
\usepackage{amsmath,amsfonts}

\newcommand{\Nom}{N_\omega}
\newcommand{\NT}{N_T}
\newcommand{\ND}{N_D}
\newcommand{\NC}{N_C}
\newcommand{\expos}{\Psi}

\newcommand{\T}{\rm T}

\theoremstyle{plain}

\theoremstyle{definition}

\theoremstyle{remark}
\newtheorem{remark}{Remark}
\usepackage{color}

\graphicspath{{Charts/}}

\begin{document}


\title{Efficient exposure computation by risk factor decomposition}

\author{C.S.L. de Graaf$^{1,2}$\footnote{Corresponding author. E-mail address: \texttt{C.S.L.deGraaf@UvA.nl}.}, 
D. Kandhai$^{1,3}$ and C. Reisinger$^{2}$ \medskip \\
\scriptsize
       $^1$ Computational Science Lab, University of Amsterdam, 1098 XH Amsterdam, The Netherlands\\
       \scriptsize
       $^2$ Mathematical Institute \& Oxford-Man Institute of Quantitative Finance, University of Oxford, \\ 
        \scriptsize
       Oxford, OX2 6GG$\vert$6ED, United Kingdom\\    
       \scriptsize    
       $^3$ Quantitative  Analytics, ING Bank, 1102 BD Amsterdam, The Netherlands
       }
\date{\today}

\maketitle

\begin{abstract}
The focus of this paper is the efficient computation of counterparty credit risk exposure  on portfolio level. 
Here, the large number of risk factors rules out traditional PDE-based techniques and allows only a relatively small number of paths for nested Monte Carlo simulations, resulting in large variances of estimators in practice.
We propose a novel approach based on Kolmogorov forward and backward PDEs, where we counter the high dimensionality by a generalisation of anchored-ANOVA decompositions. By computing only the most significant terms in the decomposition, the dimensionality is reduced effectively, such that a significant computational speed-up arises from the high accuracy of PDE schemes in low dimensions compared to Monte Carlo estimation.
Moreover, we show how this truncated decomposition can be used as control variate for the full high-dimensional model, such that any approximation errors can be corrected while a substantial variance reduction is achieved compared to the standard simulation approach.
We investigate the accuracy for a realistic portfolio of exchange options, interest rate and cross-currency swaps under a fully calibrated ten-factor model.
\end{abstract}


\section{Introduction}
Managing Counterparty Credit Risk (CCR) has become one of the core activities of financial institutions since the credit crisis in 2008. The Basel III rules introduced Credit Value Adjustment (CVA) which increased counterparty risk capital requirements \citep{Gregory2015} and has significant consequences for the valuation of derivatives. The CVA is the difference between the value of a portfolio without default risk and the true portfolio value which includes default risk of the counterparty~\citep{Pykhtin2007a}. By including CVA, one can theoretically hedge away credit risk. More recently, Debt Value Adjustment (DVA), Funding Value Adjustment (FVA) and Capital Value Adjustment (KVA) \citep{Green2014} were defined and this collection of value adjustments is now referred to as XVA. For all these adjustments, Expected Exposure (EE) and Expected Positive Exposure (EPE) are main ingredients that have to be computed from the future distribution of the underlying portfolio. In this paper, we develop an approximation method for these exposures on portfolio level.

Many traded portfolios consist of multiple underlying assets and derivatives on these assets; these are valued by models with multiple risk drivers such as stochastic interest rates, FX rates and stochastic volatility. 
Computing the future value distributions, for which closed-form solutions are typically not available, is a challenging high-dimensional computational problem. A typical benchmark for methods geared towards high-dimensional problems are basket options, where the payoff depends on a portfolio of assets, such as stocks, stock indices or currencies. These options serve the purpose of diversification and are therefore popular among investors~\citep{bouzoubaa2010}. For the valuation of these options, contributions have been made by, e.g., \citet{Jain2012} for Monte Carlo estimation under early exercise and \citet{Reisinger2012} for PDE approximations.

In the context of CVA, netting needs to be taken into account and possible negative or positive parts of the exposure distribution are to be considered. This makes CVA estimation for simple plain vanilla instruments similar to the valuation of a series of basket options with different maturity, but is more complex if the instruments themselves are more complex, i.e., no closed-from or semi-closed-form pricing formulae are available. The computational problem in that case can be formulated as the estimation of nested conditional expectations, where the inner expectations are future derivative prices conditional on the underlying risk factors, and the outer expectations, taken over the risk factors, are the expected exposures for the derivative portfolio.

For computing these risk measures, typically, Monte Carlo methods are used to sample the future states of the underlying risk factors by discretization and simulation. Portfolio values for all these states can then be computed in multiple ways, including: fully nested Monte Carlo simulation \citep{gordy2010nested, broadie2011efficient}, where derivative values along those paths are estimated by an `inner' Monte Carlo simulation; regression based approaches, introduced for American style options in \citet{Longstaff2001} and applied in the context of exposure in \citet{Karlsson2014, joshi2016least}; or a grid or Fourier based method  for the inner
expectations and Monte Carlo sampling of the outer expectation
\citep{deGraaf2014}.

All these different methods have their particular strengths and weaknesses. The na\"{\i}ve nested Monte Carlo method is generically applicable but computationally extremely expensive with a complexity of $O(\epsilon^{-4})$ for root-mean-square error (RMSE) $\epsilon$.
This can be improved when the computational budget between inner and outer simulations is optimally balanced, e.g.\ $O(\epsilon^{-3})$ when trading off `inner' bias and `outer' variance \citep{gordy2010nested}, and $O(\epsilon^{-5/2-\delta})$ for arbitrary $\delta>0$ for certain non-uniform estimators \citep{broadie2011efficient}.
Regression based approaches have
in a sense optimal complexity $O(\epsilon^{-2})$ as no extra `inner' paths are required \citep{broadie2015risk}, but 
an additional bias creeps in when leaving out sets of regression variables in higher dimensions and,
in our experience, regression methods still exhibit relatively slow convergence in applications.
In practical portfolio CVA computations, only about $10^3$ to $10^4$ paths are feasible, giving a high variance of estimators for both of these approaches, in particular for the sensitivities with respect to market factors.
PDE schemes, on the other hand, are very efficient for low dimensions but suffer from the `curse of dimensionality' -- a term used to describe the computational complexity which increases exponentially with the number of risk factors -- which made Monte Carlo methods 
the industry standard technique for problems with dimensions larger than two or three.


Notwithstanding this, we propose a method based on PDEs not just for the inner, but also for the outer expectation. We address the curse of dimensionality by
decomposition of the conditional expectations into groups of risk factors, hence breaking the high-dimensional problem up into a sequence of lower-dimensional problems, whose solutions are then assembled into a truncated decomposition.
For these low-dimensional problems 
numerical PDE methods are available which give sufficiently accurate approximations in acceptable time. 
We will find that for practically sufficient accuracy, the approximation requires solving only one one-dimensional base approximation and multiple two- and three-dimensional approximations which are used as corrections.
Perhaps practically most relevant, the approximated portfolio dynamics underlying these expansions can be used to construct control variates for Monte Carlo estimators. As such, we do not see this methodology to compete with the above simulation approaches, but to be used in conjunction with them, as it provides a way to combine seamlessly the high accuracy of PDE schemes in low dimensions and the robustness of Monte Carlo schemes in high dimensions.
This includes improvements of Monte Carlo schemes using quasi-Monte Carlo sampling (both for regression and nested simulation).


Note that 
our approach is distinctly different from dimension and variance reduction methods based on conditional sampling as in \citet{dang2015dimension, cozma2015mixed, dang2017dimension}.

The method proposed here draws on a variety of related previous work, but requires substantial extensions for the present context:
\begin{itemize}
\item
in terms of application and problem structure, the method extends \citet{ReisngerWittum2005} from basket options under Black-Scholes to portfolio risk management under a general class of models;
\item
specifically, for the exposure estimation problem formulated as nested conditional expectations, we solve a combination of forward and backward Kolmogorov equations and carry out numerical integration of the solutions;
\item
compared to the dimension-wise integration method of \citet{griebel2010dimension}, the future distribution here is unknown, such that forward PDEs have to be approximated; 
\item
compared to \citet{ReisngerWittum2005, Reisinger2012}, we
use original variables as factors instead of principal components; this makes the results more easily interpretable;
\item
compared to \citet{reisinger2015error} which provides error bounds for the constant coefficient setting, we deal with the case of non-linear SDEs and variable-coefficient PDEs,
which requires a generalised definition of the expansion;
\item
instead of theoretical error bounds, we assess the method on a complex real-world example, as outlined in the following.
\end{itemize}

For the numerical illustration, we consider typical FX portfolios consisting of multiple financial derivatives, namely Interest Rate Swaps (IRS), Cross Currency Basis Swaps (CCYS) and FX options. The risk factors are driven by a multi-dimensional Black-Scholes-2-Hull-White (BS2HW) model which captures stochasticity in the FX and the interest rates. 

In actual practical applications, one may want to consider FX and interest rate smiles and a multi-curve framework in which a stochastic basis model should be used.
In terms of the latter, however,
little is known on how to calibrate the basis vol in an implied measure, and therefore in most pricing models basis is considered as deterministic.
In the context of CCR, the problem is even more complex since one can argue that basis is a mixture of credit and liquidity and therefore it is not clear if it should be modeled as a pure IR component or whether it is a separate risk factor strongly correlated with credit. As such, the question is linked with other modeling complexities such as wrong-way risk.

These modeling aspects are beyond the scope of the paper. We emphasise though that the
computational framework presented here is rich enough to incorporate such extensions.
We demonstrate the flexibility of the approach by giving results for an extension including stochastic volatility, namely the Heston-2-Hull-White model. For the three chosen currency pairs this leads to a ten-dimensional PDE, where we demonstrate that a relative accuracy of about 1\% can be achieved for expected positive exposure computations.

The models are calibrated to market data and for these different portfolios we compute EE and EPE and compare our results with a full scale Monte Carlo benchmark, demonstrating vastly superior computational complexity.

The outline of this paper is as follows. In the following Section \ref{sec: Problem formulation}, we describe the problem set-up.
In Section \ref{sec: Risk factor decompostion}, we explain the main approximation technique used.
(A brief description of the finite difference schemes used for the PDEs in the decomposition
is given in Appendix \ref{sec:findiff}.)
In Section \ref{sec: Case study}, we set up our case studies followed by Section \ref{sec: Results} were we present our results. In Section \ref{sec: conclusion}, we provide a discussion and concluding remarks.

\section{Problem formulation}\label{sec: Problem formulation}

\noindent

We consider the general framework of a financial market described by a $d$-dimensional stochastic process $X=(X_t^i)^{1\le i \le d}_{t\ge 0}$.
In our applications, $X$ will be given by a stochastic differential equation of the form
\begin{eqnarray}
\label{SDE}
\label{Xproc}
dX_t^i &=& \mu_i(X_t,t) \, {\rm d}t + \sigma_i(X_t,t) \, {d}W^i_t, \qquad i=1,\ldots, d, \quad t>0, \\
X_0^i &=& a_i, \qquad\qquad\qquad\qquad\qquad\qquad\quad i=1,\ldots, d,
\label{XprocIC}
\end{eqnarray}
where $W$ is a $d$-dimensional standard Brownian motion with a given correlation matrix $(\rho_{i,j})_{1\le i,j \le d}$, and $a \in \mathbb{R}^d$ a given initial state,
$\mu_i$ is the drift, $\sigma_i$ the volatility.
We assume that the processes are written 
under a risk-neutral measure $\mathbb{Q}$, which is relevant for derivative valuation as well as regulatory CVA computations.


We now consider derivatives on $X$. For simplicity, we focus on the European case here.
Let the payoff function $\phi_k$ determine the amount $\phi_k({X}_{T_k})$ to be received by the holder at time $T_k$ for $1\le k\le \ND$.
The arbitrage-free 
value $V_k$ of this claim for $0\le t\le T$ is then
\[
V(x,t) = 
\mathbb{E}[\exp(-{\scriptstyle \int_t^T} \rho(X_s) \, ds) \phi(X_T) \vert X_t = x],
\]
%
%
where $\rho$ is a discount factor (and we have dropped the subscripts $k$ for simplicity), and satisfies the Kolmogorov backward PDE (see, e.g.\ \citet{Musiela2005})
\begin{eqnarray}
\label{eq: Backward Kolmogorov PDE}
\frac{\partial V}{\partial t}+\sum_{i=1}^d\mu_i\frac{\partial V}{\partial x_i}+\frac{1}{2}\sum_{i,j=1}^d\sigma_i\sigma_j\rho_ {i,j}\frac{\partial^2V}{\partial x_i\partial x_j} - \rho V&=0,\\
V({x},T)&=\phi({x}). 
\nonumber
\end{eqnarray}

For Counterparty Credit Risk (CCR), not only the time $0$ value is important, but also the evolution of this value after the deal has been entered. In this work, we use several measures to quantify CCR.
Let
\[
\sum_{k=1}^{\ND} w_k(t) V_k(X_t,t)
\]
be the value of a portfolio of ${\ND}$ derivatives at time $t$ that a financial institution holds, where $w_k$, $1\le k \le \ND$, are derivative portfolio weights.
Here, we set $w_k(t)=0$ for $t>T_k$, the expiry of the $k$-th derivative. 

First, following \citet{Pykhtin2007a}, Expected Exposure (EE) is defined as the expected value of the portfolio at time $t\geq 0$\footnote{Here, we do not include collateralisation. This can be added by including a collateral model on top of the exposure dynamics.},
\begin{eqnarray}
\text{EE}(t) = \mathbb{E} \left[ \sum_{k=1}^{\ND} w_k(t) V_k(X_t,t) \Bigg| \mathcal{F}_{0} \right], \label{eq: Expected Exposure}
\end{eqnarray}
where $\mathcal{F}_{0}$ is the filtration at time $t=0$. 
Similarly, the Expected Positive Exposure (EPE) at a future time $t<T:=\max_k T_k$ is given by
    \begin{eqnarray}
        \text{EPE}(t)=\mathbb{E}\left[\max\Bigg(0,\sum_{k=1}^{\ND} w_k(t) V_k(X_t,t)\Bigg)\Bigg|\mathcal{F}_{0}\right]\label{eq: Expected Positive Exposure}.
    \end{eqnarray}
Note that the Expected Negative Exposure (ENE) can be computed by taking the minimum instead of the maximum in equation (\ref{eq: Expected Positive Exposure}), or from ENE $=$ EE $-$ EPE. In short, EPE (ENE) gives the expected loss of the buyer (seller) of the portfolio in case he or she is in the money and the counterparty defaults, which makes them of prime importance when computing CVA and other XVAs. For example, CVA is defined as
\begin{eqnarray}
\label{Chp2: CVAeq}
 \mathrm{CVA}&=& (1-R) \int_0^T \mathbb{E}\left[\exp(-{\scriptstyle \int_t^T} \rho(X_s) \, ds)\max\Bigg(0,\sum_{k=1}^{\ND} w_k(t) V_k(X_t,t)\Bigg)\Bigg|\mathcal{F}_{0}\right] d\text{PD}(t),\\
 &=& (1-R) \int_0^T  D(t) \, \text{EPE}^*(t) \, d\text{PD}(t),\nonumber
\end{eqnarray}
where $R$ is the recovery rate, $\text{EPE}^*(t)$ is the discounted EPE, and $\text{PD}(t)$ denotes the default probability of the counterparty at time $t$~\citep{Gregory2010}.  Note that (\ref{Chp2: CVAeq}) incorporates the debatable but common assumption that the portfolio value is independent of default events (no `` wrong-way risk'').

In this paper, we focus solely on the computation of EE and EPE.
%
%
%
For exposure calculations at time 0,
we then consider functionals of the form
\begin{eqnarray}
\nonumber
V(x,0;t) &=& \mathbb{E}\Bigg[\expos\Bigg(\sum_{k=1}^{\ND} w_k(t) V_k(X_t,t)\Bigg) \Bigg \vert X_0 = x \Bigg] \\
&=& \mathbb{E}\Bigg[\expos\Bigg(\sum_{k=1}^{\ND} w_k(t) \mathbb{E}\Big[\phi_k(X_{T_k})\vert {X}_t \Big] \Bigg) \Bigg  \vert X_0 = x \Big],
\label{expect}
\end{eqnarray}
where $\expos$ is an exposure function, e.g., $\expos(x) = x$ in the case of EE, or $\expos(x) = \max(x,0)$ for EPE.
This can be adapted in the obvious way for American or path-dependent options.

\begin{remark}
\label{rem:fwdbwd}
To find $V(x,0;t)$, the expected (positive) exposure at time $t$ as seen from time $0$, we could therefore solve backward PDEs in $(t,T)$
for $V_k(\cdot,t)$, and then another backward PDE over $(0,t)$ with $\expos[\sum_{k=1}^{\ND} w_k(t) V_k(\cdot,t)]$ as terminal condition.
This would however require the solution of a different backward PDE in the second stage for different $t$.
To avoid this, in the following we utilise a forward PDE for the transition densities to compute exposures.

The reason we use the backward PDE to approximate $V_k$ instead of using the density function, is that we need the value of derivatives in the whole state space reachable by $X$. 
\end{remark}

Let  
$p({x},t;{a},0)$ be the transition density function of $X_t$ at $x$ given state ${a}$ at $t=0$,
 then 
\begin{eqnarray}
\label{integral}
V(x,0;t) = \int_{\mathbb{R}} p(y,t;x,0) \expos\Bigg(\sum_{k=1}^{\ND} w_k(t) V_k(y,t)\Bigg) \, {d}y.
\end{eqnarray}
Here, $p$ is given through an adjoint relation by the Kolmogorov forward equation
\begin{eqnarray}
\label{eq: Forward Kolmogorov PDE}
-\frac{\partial p}{\partial t}-\sum_{i=1}^d\frac{\partial}{\partial x_i}\left(\mu_i p\right)+\frac{1}{2}\sum_{i,j=1}^d\frac{\partial^2}{\partial x_i\partial x_j}\left(\sigma_i\sigma_j\rho_ {i,j}p\right)=&0,\hspace{2cm}\\
p(x,0;a,0)=&\delta(x-a),\hspace{1cm}\label{eq: Forward Kolmogorov BC}
\end{eqnarray}
where $\delta$ is the Dirac distribution centred at $0$. 

Hence, we can obtain expected (positive) exposures by the solution of one forward PDE for the density and one backward PDE for each derivative in the portfolio, plus one integration for each $t$. 


\section{
Approximation by risk factor decomposition}
\label{sec: Risk factor decompostion}

The principal difficulty in practice arises from the dimensionality $d$ of $X$ in (\ref{Xproc}). 
Although $\ND$ in (\ref{expect}) is typically very large, the computational complexity is linear in $\ND$, while it is exponential in $d$.
In this section, we introduce an approximation technique which makes large $d$ computationally manageable.

To introduce the concepts, we focus in Section \ref{subsec-anova} on the problem of approximating 
\begin{eqnarray}
\label{expect1}
V(x,t) &=& \mathbb{E}[\phi(X_T) \vert X_t=x] \\
&=& \int_{\mathbb{R}^d} \phi(y) \ p(y,T;x,t) \, {d}y,
\label{integral2}
\end{eqnarray}
i.e.\ for simplicity the discount factor is 1, where $p(\cdot,T;x,t)$ is the probability density function of $X_T$. 
We first discuss an extension of the anchored-ANOVA concept to PDEs with variable coefficients and the use of a ``moving anchor''.
In Section \ref{subsec:cv} we explain how the approximations can be used as effective control variates for unbiased Monte Carlo estimators.
The application to the complete expected exposure problem (\ref{integral}) will be discussed in Section \ref{subsec:portfolios}.

\subsection{An anchored-ANOVA-type approximation}
\label{subsec-anova}

The proposed method extends anchored-ANOVA decompositions as considered by \citet{griebel2010dimension}
in the context of integration problems. In the setting and notation of (\ref{integral2}), the methodology in \citet{griebel2010dimension}
chooses an \emph{anchor} $a \in \mathbb{R}^d$ and then, for a given index set $u\subset \mathcal{D} = \{i \, : \, 1\le i\le d\}$,
defines projections of the integrand $f=\phi p$ by $f(a\backslash x_u)$,
where $a\backslash x_u$ denotes a $d$-vector such that
\begin{eqnarray}
\label{ax}
(a\backslash x_u)_i = \left\{
\begin{array}{rl}
x_i & i\in u, \\
a_i & i\notin u.
\end{array}
\right.
\end{eqnarray}
This leads to a decomposition of $f$ into lower-dimensional functions, which can be exploited for successive quadrature approximations (see  \citet{griebel2010dimension} and the references therein).

In contrast to there, we cannot assume here that the joint probability density function for $X_T$ is analytically known and therefore we have to 
consider the dynamics of $X$ directly.
To improve the accuracy, e.g.\ to account for the term-structure of interest rates, here we do not concentrate the measure at  a fixed anchor $a$, 
but we allow the anchor to move with a certain conditional expectation of the underlying process, as detailed in the following.

We first define a deterministic approximation to $X$ from (\ref{Xproc}), (\ref{XprocIC}) by
\begin{eqnarray*}
\xi^i(t) &=& \mathbb{E}\left[\overline{X}_t^i \left\vert \overline{X}^i_0 = a_i \right. \right], \qquad\qquad\qquad\qquad\qquad\qquad\qquad 1\le i \le d, \;\; t\ge 0, \\
\overline{X}_t^i  
&=& a_i + \int_{0}^t \mu_i(a_1,\ldots, a_{i-1}, \overline{X}^i_s, a_{i+1},\ldots, a_d  ,s) \, {d}s + \\
&& \hspace{2 cm} + \int_{0}^t \sigma_i(a_1,\ldots, a_{i-1}, \overline{X}^i_s, a_{i+1},\ldots, a_d,s) \, {d}W^i_s, \qquad t\ge 0.
\end{eqnarray*}
In \citet{Reisinger2012}, a more simplistic approach is used where the drift is first approximated by $\mu(a,0)$, and then eliminated by a coordinate transformation $x \rightarrow x- \mu(a,0) \!\ t$. The above construction takes into account more information about the term structure and also deals systematically with non-constant volatilities.

For given subset $u \subseteq \mathcal{D}$, we then define a process
\begin{eqnarray}
\label{eq:anchor}
\label{eq:anova}
X^u_t = \left\{
\begin{array}{rl}
a_i + \int_{0}^t \mu_i(\xi(t)\backslash X^u_s,s) \, {d}s + \int_{0}^t \sigma_i(\xi(t)\backslash X^u_s,s) \, {d}W^i_s, & i \in u, \\
\xi^i(t),
& i \not\in u,
\end{array}
\right.
\end{eqnarray}
where $\xi(t)\backslash X^u_s$ follows the notation from (\ref{ax}).

Here, we have replaced a subset of the processes by their conditional expectations. 
For instance, if $X^1$ is an exchange rate and $X^2$ and $X^3$ are the domestic and foreign short rate in a Hull-White model, then $\xi^1$ is the expectation of the exchange rate under a constant interest rate model, while $\xi^2$ is simply the expectation of the domestic short rate.
The point is that the dynamics is effectively of dimension $|v|$, and the expectation (\ref{expect1}) can be approximated by lower-dimensional problems.

%
%
Accordingly, we define
\begin{eqnarray}
\label{eq:backward-v}
{V}_u(a;x,t) = \mathbb{E}[\phi(X_T^u) \vert X^u_t = x],
\end{eqnarray}
where the right-hand side depends on $a$ implicitly through the definition of $X^u_t$ in (\ref{eq:anova}).

Specifically, the backward PDE for (\ref{eq:backward-v}) under (\ref{eq:anchor}) is
\begin{eqnarray}
\label{eq: Backward Kolmogorov PDE-move}
\frac{\partial V_u}{\partial t}+
\underbrace{
\sum_{i\in u} \mu_i(\xi(t)\backslash x_u) \frac{\partial V_u}{\partial x_i}+
\frac{1}{2} \sum_{i,j \in u}\sigma_i(\xi(t)\backslash x_u) \sigma_j(\xi(t)\backslash x_u) 
\rho_ {i,j}\frac{\partial^2V_u}{\partial x_i\partial x_j}}_{\equiv \ \mathcal{L}_u V_u}
&=0,  \\
V_u({x},T)&=\phi({x}).
\label{eq: Backward Kolmogorov BC-move}
\end{eqnarray}
The significance of the arguments $\xi(t)\backslash x_u$ in the coefficients of (\ref{eq: Backward Kolmogorov PDE-move}) is that the full information on the coordinates $x_j$, $j\in u$ is used, while the solution $V(a;a,0)$ at the anchor point can be computed solving only a $|u|$-dimensional PDE in the variables $x_j$, $j\in u$.


\begin{remark}
\label{rem:disc}
The generalisation from (\ref{eq:backward-v}) to
\[
{V}_u(a;x,t) = \mathbb{E}[\exp(-{\scriptstyle \int_t^T} \rho(X_s^u) \, ds) \phi(X_T^u) \vert X^u_t = x]
\]
is straightforward and the corresponding backward PDE instead of (\ref{eq: Backward Kolmogorov PDE-move}) is
\[
\frac{\partial V_u}{\partial t} + \mathcal{L}_u V_u + \rho(\xi(t)\backslash x_u) V_u = 0.
\]
\end{remark}

For future reference, we also have the forward PDE
\begin{eqnarray}
\label{eq:forward-v-move}
\frac{\partial p_u}{\partial t}+\sum_{i\in u}\!\frac{\partial}{\partial x_i}\!\left(\mu_i(\xi(t) \backslash x_u) p_u\right)-
\!\frac{1}{2}\!\sum_{i,j\in u}\!\!\frac{\partial^2}{\partial x_i\partial x_j}\!\left(\sigma_i(\xi(t) \backslash x_u)  \sigma_j(\xi(t)\backslash x_u) \rho_ {i,j} p_u\right)=&0,\hspace{1cm}\\
p_u(x,0;a,0)=&\delta(x-a).\label{eq:forward-v-ic-move}
\end{eqnarray}

%

%
%
%
%
%

From here onwards, we can follow the path of \citet{griebel2010dimension} and the references therein, to
define a decomposition recursively through a difference operator $\Delta$, by $\Delta {V}_{\emptyset} = V_{\emptyset}$ and, for $u\neq \emptyset$,
\begin{eqnarray}
\label{surplus}
\Delta {V}_u(a;x,t) &=&{V}_u - \sum_{w\subset u} \Delta {V}_w
\;\; = \;\; \sum_{w\subseteq u} (-1)^{|w|-|u|} {V}_w,
\end{eqnarray}
where the inclusion in the first summation is strict.
This is indeed a decomposition because 
\begin{eqnarray}
\label{anova}
V(a;x,t) &=&  \sum_{u\subseteq \{1,\ldots,d\}} \Delta {V}_u
\;\; =\;\; \sum_{k=0}^d \sum_{|u|=k} \Delta {V}_u.
\end{eqnarray}

We note that for $u=\{i_1,\ldots,i_{|u|}\}$, $\Delta {V}_u(a; \xi(t)\backslash x_u,t)$ only depends non-trivially on the sub-set of coordinates $\{x_{i_1},\ldots, x_{i_{|u|}}\}$, and satisfies (\ref{eq: Backward Kolmogorov PDE-move}).
 Therefore, ${V}_u$ and hence $\Delta {V}_u$ can be found by the solution of problems of dimension not higher than $|u|$.

As an example, consider $d=3$ and $u=\{1,2\}$, then
\begin{eqnarray*}
\label{surplusex}
\Delta {V}_u(a;x,t) &=& {V}_{\{1,2\}} - \left(\Delta {V}_{\{1\}} + \Delta {V}_{\{2\}} + \Delta {V}_{\emptyset} \right)\\
&=&  {V}_{\{1,2\}} - {V}_{\{1\}} - {V}_{\{2\}} + {V}_{\emptyset}.
\end{eqnarray*}

We can now define an approximation by 
%
\begin{eqnarray}
\label{truncated}
V_{0,s}(a;x,t) &=&  \sum_{k=0}^s \sum_{|u|=k} \Delta {V}_u
\;\; =\;\;  \sum_{k=0}^s c_k \sum_{|u|=k}  V_u,
\end{eqnarray}
where $c_k$ are integer constants, which also depend on $s$ and $d$ (we suppress this to keep the notation simple).
The point is that the approximation $V_{0,s}(a;a,t)$ at the anchor can be found by solving PDEs of dimension at most $s$.

\begin{remark}
The approximations $V_{0,1}$ and $V_{0,2}$ have certain similarities with delta and delta-gamma approximations, respectively (see, e.g.,
\citet{alexander2006minimizing}).
In the latter, the derivative value is approximated by
\[
V(X_t,t) \approx V(a,0) + 
\frac{\partial V}{\partial t} \ t
+ (X_t-a)^T \ \nabla V
+ \frac{1}{2}(X_t-a)^T \ \nabla \nabla^T V (X_t-a),
\]
where $a=X_0$, $\nabla V$ is the gradient or delta and $\nabla \nabla^T V$ the Hessian or gamma. One notable difference is that
$V_{0,2}$ does not make any approximations if $V$ is two-dimensional or a sum of two-dimensional functions, while delta-gamma always involves linear and quadratic approximations.
\end{remark}


An alternative approximation, extending (\ref{truncated}), is given by
\begin{eqnarray}
\label{truncated_rs}
V_{r,s}(a;x,t) &=&  \sum_{k=0}^s c_k \sum_{|u|=k} V_{u\cup {\scriptscriptstyle \{1,\ldots,r\}}}, \quad r+s\le d.
\end{eqnarray}
Here, we always retain the first $r$ coordinates, and apply the splitting only to the remaining $d-r$ coordinates.
The  approximation $V_{r,s}(a;a,t)$ at the anchor can be found by solving PDEs of dimension at most $r
+s$.
It is this approximation 
that we will use in the numerical computations later on in the paper, with $r=1$ and $s=1$ or $s=2$. The coordinate $x_1$ is chosen to capture most of the dynamics, either through prior knowledge or small pilot runs with reduced accuracy.

For instance, for $s=1$, $r=1$, we have
\begin{eqnarray}
\label{truncatedex1}
V_{1,1}(a;x,t) &=& \sum_{1<i\le d} \left( V_{\{1,i\}} \!-\! V_{\{1\}}\right) \, + \, V_{\{1\}} \\
&=& \sum_{1<i\le d} V_{\{1,i\}} \; - (d-2) \; V_{\{1\}},
\nonumber
\end{eqnarray}
i.e., $c_0=-(d-2)$, $c_1=1$ in (\ref{truncated_rs}), and, for $s=2$, $r=1$,
\begin{eqnarray}
\label{truncatedex}
\!\!\!\! V_{1,2}(a;x,t) \!\! &=&  \!\!\!\! \sum_{1<i< j\le d} \left(V_{\{1,i,j\}} \! -\!  V_{\{1,i\}} \!-\! V_{\{1,j\}} + V_{\{1\}}\right)
\, + \sum_{1<i\le d} \left( V_{\{1,i\}} \!-\! V_{\{1\}}\right) \, + \, V_{\{1\}} \\
\!\! &=& \!\!\!\! \sum_{1<i< j \le d} V_{\{1,i,j\}} \;-\; (d-2) \sum_{1<i\le d} V_{\{1,i\}} \; + \frac{(d-2)(d-1)}{2} \; V_{\{1\}},
\nonumber
\end{eqnarray}
i.e., $c_0 = \binom{d-1}{2}$, $c_1 = -(d-2)$, $c_2=1$.

%
%
%
%


\subsection{A control variate}
\label{subsec:cv}

In some situations, the approximation (\ref{truncated_rs}) with small $r$ and $s$ will be sufficiently accurate. There is an expectation -- but no guarantee in general -- that increasing either $r$ or $s$ will improve the accuracy. The computation of further terms may also be prohibitively expensive due to the high dimensionality and number of PDEs involved. In such cases, it will be valuable to simulate certain corrections by a MC method -- instead of computing them by PDEs -- to obtain more accurate approximations.


Therefore, we now discuss a way to turn the approximation (\ref{truncated_rs}) into a control variate for a Monte Carlo scheme for (\ref{expect1}).
Denote by $\{\omega_k: 1 \le k\le {\Nom}\}$ a set of ${\Nom}$ independent samples of the $d$-dimensional Brownian motion $W$, and $X_T({\omega_k})$ the (strong) solution to (\ref{SDE}) for a given sample path ${\omega_k}$ of $W$.

The standard estimator of $V(x,0)$ in (\ref{expect1}) is then
\begin{eqnarray*}
\widehat{V}^{\Nom} = \frac{1}{{\Nom}} \sum_{k=1}^{\Nom} \phi(X_T(\omega_k)).
\end{eqnarray*}

Denote by $\Phi = \phi(X_T)$ the random payoff, i.e.\ for event $\omega_k$ define $\Phi({\omega_k}) = \phi(X_T({\omega_k}))$ and
\[
\Phi_{r,s} = \sum_{k=0}^s c_k \sum_{|u|=k} \phi(X_T^{u\cup {\scriptscriptstyle \{1,\ldots,r\}}})
\]
(compare with (\ref{truncated_rs})), then we define the estimator
\begin{eqnarray}
\label{varredest}
\widehat{V}^{\Nom}_{r,s} 
&=&  \frac{1}{{\Nom}} \sum_{k=1}^{\Nom} \left(\Phi(\omega_k) -
\alpha \left(\Phi_{r,s}(\omega_k) - V_{r,s} \right)
\right),
\end{eqnarray}
where $\alpha$ will be determined later to achieve the best variance reduction, and $V_{r,s}$ given by (\ref{truncated_rs})
will be constructed from numerical PDE solutions.
From
\begin{eqnarray*}
V &=& \mathbb{E}[\phi(X_T)], \qquad
V_{u\cup {\scriptscriptstyle \{1,\ldots,r\}}} \;\;=\;\; \mathbb{E}[\phi(X_T^{u\cup {\scriptscriptstyle \{1,\ldots,r\}}})],
\end{eqnarray*}
we get
\begin{eqnarray*}
\mathbb{E}[\widehat{V}^{\Nom}_{r,s}] &=& V,
\end{eqnarray*}
i.e., the standard estimator and the control variate estimator are both unbiased.

Typically, we will find that $\alpha\approx 1$ is optimal. Indeed, if $\alpha=1$, then
\begin{eqnarray}
\label{varredest1}
\widehat{V}^{\Nom}_{r,s} &=& V_{r,s} \, + \, \frac{1}{{\Nom}} \sum_{k=1}^{\Nom} \left( \Phi(\omega_k) - \Phi_{r,s}(\omega_k)  \right),
\end{eqnarray}
i.e.\ we sample the approximation error of the truncated decomposition.
The motivation for choosing $\widehat{V}^{\Nom}_{r,s}$ over the standard estimator $\widehat{V}^{\Nom}$ is that for given ${\omega_k}$, 
$ \Phi_{r,s}$ and $\Phi$ will be close and therefore the variance much reduced compared to the standard estimator.

The value of $\alpha$ which minimises the variance is determined by the co-variances of the control variate and the standard estimator, which can be approximated by the estimators (see \citet{glasserman2004})
\begin{eqnarray*}
\widehat{\sigma^2} = \frac{1}{{\Nom}}\sum_{k=1}^{{\Nom}}\left(\Phi_{r,s}(\omega_k)-V_{r,s})\right)^2, \qquad
\widehat{\rho} = \frac{1}{{\Nom}}\sum_{k=1}^{{\Nom}}\left(\Phi(\omega_k)-\widehat{V}^{\Nom}\right)\left(\Phi_{r,s}(\omega_k)-V_{r,s}\right),
\end{eqnarray*} 
and the optimal value is estimated as $\widehat{\alpha} = {\widehat{\rho}}/{\widehat{\sigma^2}}$.
The variance reduction is then approximately
\begin{eqnarray*}
\frac{\mathrm{Var}\left[\Phi - \alpha \left(\Phi_{r,s} - V_{r,s} \right)\right]}{\mathrm{Var}\left[\Phi \right]} \;\; \approx \;\; 1-
\frac{\mathrm{cov}\left(\Phi,\Phi_{r,s}\right)}{\sqrt{\mathrm{var}\left(\Phi\right)} \sqrt{\mathrm{var}\left(\Phi_{r,s}\right)}}.
\end{eqnarray*}

Summarising, in situations where the approximation $V_{r,s}$ from the previous section is not accurate enough, it can be corrected with a 
relatively small number of Monte Carlo samples of the correction terms (the right-hand sum) in (\ref{varredest1}).

\subsection{Application to derivative portfolios}
\label{subsec:portfolios}


We now discuss approximations to expected exposures, as per (\ref{expect}). 
To benefit from the dimension reduction of risk-factor decomposition, we compute the solutions of all PDEs involved in the spirit
of Section \ref{subsec-anova}.

The general principle is to replace the process $X$ in (\ref{expect}) by $X^v$ as defined in 
(\ref{eq:anchor}).
Thus, we approximate $V$ in (\ref{integral}) by
\begin{eqnarray}
\label{integral-v} V_u(a;x,t) = \int_{\mathbb{R}} p_u(y,t; x,0) \, \expos\Bigg(\sum_{k=1}^{\ND} \omega_k(t) V_{k,u}(\xi(t) \backslash y_u ,t)\Bigg) \, {d}y,
\end{eqnarray}
where $p_u$ is the transition density function of $X^u$, 
which satisfies the forward PDE (\ref{eq:forward-v-move}), (\ref{eq:forward-v-ic-move})
instead of (\ref{eq: Forward Kolmogorov PDE}).
Similarly, $V_{k,u}$ is given as the solution to (\ref{eq: Backward Kolmogorov PDE-move}), with payoff function $\phi_k$ in (\ref{eq: Backward Kolmogorov BC-move}).
The key point is that we compute $p_u$ and $V_{k,u}$ by solving $|u|$-dimensional PDEs, and (\ref{integral-v}) is a $|u|$-dimensional integration problem, as $p$ is a Dirac measure in dimensions $\mathcal{D}\backslash u$.

Then $V_{r,s}$ is defined  by (\ref{integral-v}) and (\ref{truncated_rs}).
The complexity of the whole computation is linear in $\ND$, and exponential in $r+s$.

For products where model derivative prices are available in closed form (such as most swaps), we will use those in the computations for speed and accuracy.
For others (such as options), we will use numerical PDE approximations as outlined in Appendix \ref{sec:findiff}.

\section{Case studies}\label{sec: Case study}

In this section, we present the market set-up for the detailed numerical studies in Section \ref{sec: Results}, where we compute expected exposures (\ref{eq: Expected Exposure}) and expected positive exposures (\ref{eq: Expected Positive Exposure}) of portfolios with increasing complexity.

The example consists of various exchange rate and interest rate products, and is chosen to be representative for the scenario where each product in a portfolio is exposed to one or more of a pool of risk factors. So for instance an exchange option of medium term maturity is exposed to interest rate risk in both the domestic and foreign markets, and conversely the domestic interest rate affects an exchange option as well as, say, a domestic zero coupon bond. 


\subsection{Driving risk factors}
\label{subsec:drivers}

We consider a portfolio of FX  and interest rate products. Each FX rate is assumed to be governed by a full Black-Scholes-2-Hull-White (BS2HW) model (see, e.g.\ \citet{clark2011foreign}).
 Every FX rate $F^i_t$ thus has a stochastic domestic and foreign interest rate.
For clarity of exposition, we consider the case where all exchange rates are relative to a single currency, in the case below the EUR, so that
we can write the joint dynamics of FX and interest rates as
\begin{subequations}
\label{FRprocessesBS}
    \begin{eqnarray}
         dF^i_t   =& (R_t^{\text{d}}-R_t^{\text{f},i}) F^i_t \, dt + \sigma^i(t) F^i_t \, dW^{{\rm F},i}_t, \,\;\qquad\qquad\qquad \qquad \qquad \qquad 1\le i \le m, \label{eq:SDE BS}\\
         dR^{\text{d}}_t   =& \lambda_{\text{d}}(\Theta_{\text{d}}(t)-{R}^{\text{d}}_t  ) \, dt + \eta_{\text{d}} \, dW^{\rm d}_t, \qquad \qquad \qquad \qquad\qquad\qquad \qquad \qquad \qquad
         \label{eq:SDE r_d}\\
         dR^{\text{f},i}_t   =& \left[\lambda_{\text{f}}^i (\Theta_{\text{f}}^i(t)-R^{\text{f},i}_t  )-\eta^i_{\text{f}} \rho_{I(i), J(i)} 
         \sigma^i(t)\right] \, dt + \eta_{\text{f}}^i \, dW^{{\rm f},i}_t,  
         \qquad \qquad \qquad \quad 1\le i\le m, \label{eq:SDE r_f}    
         \end{eqnarray}
\end{subequations}
with given Brownian motions $(W^i)_{1\le i\le 2m+1}=(W^{{\rm F},1},W^{\rm d},W^{{\rm f},1},W^{{\rm F},2},W^{{\rm f},2},W^{{\rm F},3},W^{{\rm f},3})$, with 
\begin{eqnarray}
         d[W^i,W^j]_t &= \rho_{i,j} \, dt, \qquad\qquad\qquad\qquad\qquad\qquad\qquad\qquad \mbox{ for } 1\le i, j \le 2 m+1, \label{eq:dW_is}
\end{eqnarray}
and $I(i)$ is the index of the $i$-th exchange rate and $J(i)$ the index of the $i$-th foreign rate (e.g., $I(1)=1$, $J(1)=3$).
Here, $R^{\text{d}}$ is the domestic short rate and $R^{\text{f},i}$ the foreign short rate for exchange rate $F^i$, $i=1,\ldots,m$, with $m+1$ the total number of markets.

The interest rates are modeled by a mean-reverting Hull-White process with $\Theta_{\text d}(t)$ and $\Theta_{\text f}^i(t)$, $1\le i\le m$, designed to fit the forward rate curve in the respective markets.
More details on the calibration will be given in Section \ref{subsec:calibration}.

We defer the discussion of
stochastic volatility models 
to Section \ref{subsec:Stochastic Volatility}, but remark here that it is generally
not straightforward to extend these models from a single currency pair to multiple pairs while preserving important FX characteristics like symmetry and the triangle inequality (see \citet{Doust2012, DeCol2013}).

In the numerical examples, we choose EURUSD, EURGBP and EURJPY, i.e.\ $m=3$. Together with piecewise constant volatility, the interest rates in the EUR (domestic) and USD, GBP and JPY (all foreign) markets, this gives $d=2m+1 = 7$ risk factors. In the case of stochastic volatility, EURUSD, EURGBP and EURJPY are all modelled with a Heston model such that there are 10 risk factors.
This example is complex enough in the sense that a 7- or 10-dimensional PDE solution is not feasible and that we can demonstrate the effect of different decompositions.




\subsection{Derivatives}\label{subsec: Derivatives}


We consider portfolios of cross-currency and interest rate swaps and FX options.

\paragraph{Zero-coupon bonds}
Although the portfolios below do not contain any bonds explicitly, the swaps studied here are all based on coupon legs which can be replicated by (and hence valued from the time $0$ value of) a string of zero-coupon bonds that pay 1 at different times $T$. 
The time $t$ value a zero-coupon bond, $P(t,T)$, under Hull-White for the short rate $R$ is 
\begin{eqnarray}
\label{zero-coupon}
P(t,T) &=& \frac{P(0,T)}{P(0,t)}e^{A(t,T)-B(t,T) R_t},
\end{eqnarray}
where
\begin{eqnarray*}
A(t,T) = B(t,T)f(t)-\frac{\eta^2}{4\lambda}B^2(t,T)(1-e^{-2\lambda t}), \qquad
B(t,T) = \frac{1-e^{-\lambda(T-t)}}{\lambda},
\end{eqnarray*}
and
\begin{eqnarray}
f(t) &=& \frac{-\log\left(P(0,t)\right)}{t}, \qquad t\geq 0, \label{eq: yield formula}
\end{eqnarray}
see \citet{Filipovic2009}.
For simplicity, we assume here a single curve framework and do not make a distinction between discounting and
forwarding curves (see the discussion at the end of the introduction).

\paragraph{Cross-currency swap (CCYS)}
We consider a series of FX forward swaps settled in arrears, with floating payment and receiving leg. 
Note that both legs need to be valued in the domestic currency, for which the future exchange rate is used.
Following \citet{Filipovic2009}, the value $V_{\rm ccy}(t,T_0,T)$ of the swap at time $t\leq T_0$ with payment dates $T_1,\ldots,T_{\NC}=T$ is
\begin{eqnarray*}
 V_{\rm ccy}(F_t,t;T_0,T) = C^{\text{d}}(t,T_0,T) - M\ \! F_{t} \ \! C^{\text{f}}(t,T_0,T),
\end{eqnarray*}
where $F_t$ is the FX rate at time $t$, $C^{\text{d}}(t,T_0,T)$ and $C^{\text{f}}(t,T_0,T)$ are the domestic and foreign floating rate notes, 
and $M$ is the moneyness as a percentage of the (future) FX rate, e.g.\ 100\% is referred to as At-The-Money (ATM), 105\% as In-The-Money (ITM) and  95\% as Out of-The-Money (OTM) . The values of the floating rate notes are
 \begin{eqnarray*}
 C^{\text{d/f}}(F_t,t;T_0,T) = P^{\text{d/f}}(t,T)\Lambda^{\text{d/f}} + \sum_{i=1}^{\NC} \left(P^{\text{d/f}}(t,T_{i-1})-P^{\text{d/f}}(t,T_i)\right)\Lambda^{\text{d/f}} = P^{\text{d/f}}(t,T_0)\Lambda^{\text{d/f}},
\end{eqnarray*}
where $\Lambda^{\rm d/f}$ is the notional in domestic or foreign currency, 
$\Lambda^{\text{f}} = \Lambda^{\text{d}}/F_0$.

\paragraph{Interest rate swap}
Here, we consider a floating versus fixed interest rate swap, where the fixed leg is defined as a fixed coupon bond, such that the value 
with notional $\Lambda$ equals
\begin{eqnarray*}
 V_{\rm irs}(t) = \left(K \Delta T \, \sum_{i=1}^{\NC} P(t,T_{i})+P(t,T) - P(t,T_0)\right)\Lambda, 
\end{eqnarray*}
where $K$ is the fixed rate and $\Delta T = T_{i}-T_{i-1}$ the interval between payment dates (see \citet{Filipovic2009}).

\paragraph{FX option} Finally, we consider an FX call option on the EURUSD FX rate with strike $K$, notional $\Lambda_{\rm Opt}$ and maturity $T$ with payoff 
\begin{eqnarray*}
\phi(F_T) = \max\left(0 , F_T - K\right)\Lambda_{\rm Opt}.
\end{eqnarray*}
The value of the option at time $t<T$, under the BS2HW model from Section \ref{subsec:drivers}, is equal to
\begin{eqnarray*}
 V_{\rm opt}(F_t, R^{\rm d}_t, R^{\rm f}_t, t) =\mathbb{E}[\exp(- {\scriptstyle \int_t^T} R_s^{\rm d} \, ds) \phi(F_T) \vert X_t],
\end{eqnarray*}
where $X_t = (F_t, R^{\rm d}_t, R^{\rm f}_t)$ follow (\ref{FRprocessesBS}).
For more details on these options we refer to \citet{Filipovic2009} and  \citet{brigo2013}. 
We only note that the option value function is the solution of the corresponding three-dimensional backward PDE; see also Remark \ref{rem:disc}.

\paragraph{Contract parameters}
The CCYS are driven by the FX rate and the foreign and domestic interest rate processes. The contract specific parameters are shown in Table \ref{Table: CCYS Parameters}. Note that the trades have different maturity, notional and moneyness. 

\begin{table}
\begin{center}
\caption{Parameters for the CCYS and IRS.}
\begin{minipage}{110mm}
{\begin{tabular}{{lcccc}}\toprule
   & &\text{EURUSD}    &\text{EURGBP} &\text{EURJPY}   \\ \midrule
Moneyness & M& 100\% & 95\% & 105\%   \\
Maturity &$T$& 5Y& 3Y&2Y \\
Notional &$\Lambda^{\rm d}$& 100& 100&50 \\
Number of coupons &${\NC}$& 100& 100& 100  \\\bottomrule
\end{tabular}}
\end{minipage}
\hspace{0.5 cm}
\begin{minipage}{30mm}
{\begin{tabular}{{cc}}\toprule
 & IR swap  \\ \midrule
$M$ & 100\%  \\
$T$& 5Y\\
$\Lambda$& 150\\
${\NC}$& 100 \\\bottomrule
\end{tabular}}
\label{Table: CCYS Parameters}
 \label{Table: IRS Parameters}
\end{minipage}
\end{center}
\end{table}

The interest rate swap is traded on the EUR interest rate, ATM, where the swap rate is such that the initial value of the trade at inception ($t=0$) is equal to zero. The specific parameters can be found in Table \ref{Table: IRS Parameters}. 

The option strike is set at $K = 0.95 F_{0}^{1}$ and the maturity to $T = 4$. The notional of this option is set to $\Lambda_{\rm Opt}=100$.

\subsection{Market and calibration}

\label{subsec:calibration}

As the behaviour and accuracy of the approximation method from Section \ref{sec: Risk factor decompostion} may depend significantly on the model parameters, we perform a careful calibration to market data before undertaking numerical tests.

\begin{figure}[ht!]
\centering
\subfigure[Time-dependent volatilities for the different FX rates, bootstrapped to the future ATM implied vols. ]{
\includegraphics[width=0.46\textwidth]{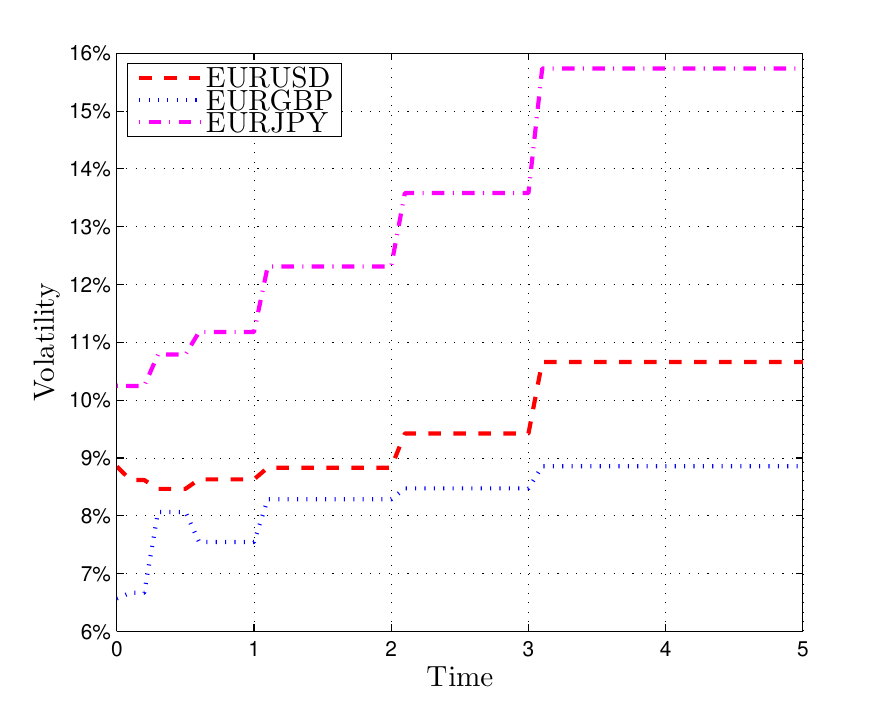}\label{subfig: Vols}
}
\hfill
\subfigure[Yields over time for the different interest rates as determined by equation (\ref{eq: yield formula}).  ]{
\includegraphics[width=0.46\textwidth]{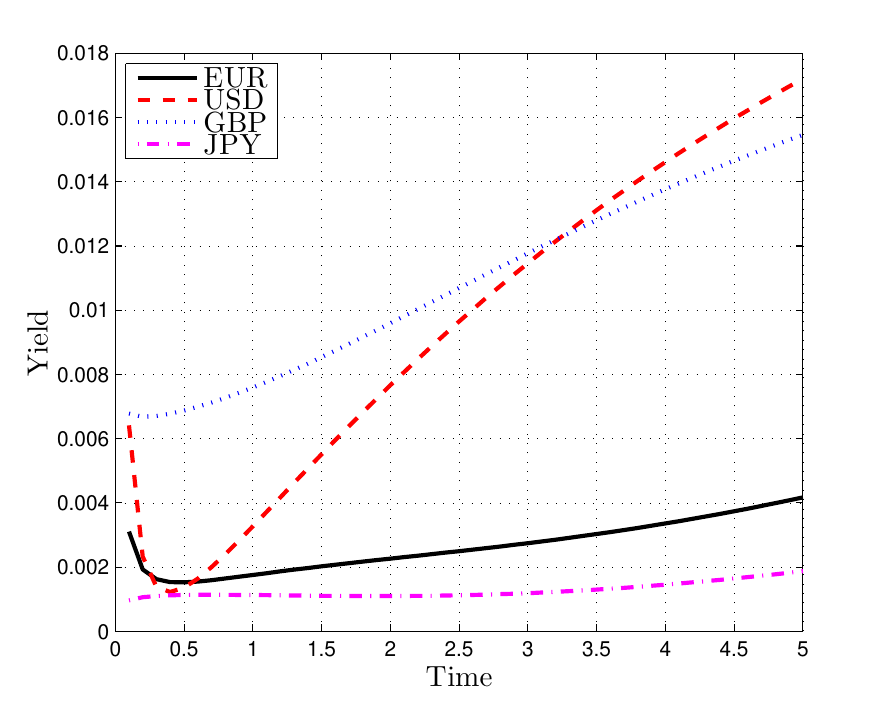}\label{subfig: Yields}
}
\caption{Volatilities and yields over time on 2 December 2014. }\label{subfig: implied vols}
\end{figure}

We use a data set from 2 December 2014. At this time, the interest rates were low, which resulted in low yields for all markets. 
Shown in Figure \ref{subfig: Yields} are the yields up to 5 years, computed by polynomial interpolation of the  market forward curve $f(t)$ as per (\ref{eq: yield formula}). They are used for bond pricing via (\ref{zero-coupon}) and
to calibrate $\Theta \in \{\Theta_{\text d}(t), \Theta_{\text f}^i(t), 1\le i \le m\}$ in (\ref{FRprocessesBS}) by
\begin{eqnarray}
\label{eq:theta}
\Theta(t) &=& \frac{1}{\lambda}\frac{{d} f(t)}{{d} t} + f(t) + \frac{\eta^2 }{2\lambda^2}(1-e^{-2\lambda t}) \label{eq: Theta}
\end{eqnarray}
with $f(t)$ from (\ref{eq: yield formula}),
which gives an exact fit to bond prices (see \citet{Filipovic2009}). 

For the FX volatilities in (\ref{eq:SDE BS}), we assume a piecewise constant volatility function which follows the ATM volatilities over time,
\begin{eqnarray*}
    \sigma(t) &= \sum_{i=1}^{N_M} \sigma_{i}1_{\{T_{i-1}<t\leq T_i\}}, 
\end{eqnarray*}
where $0=T_0<\ldots<T_{N_M}=T_{max}$ is a partition of the interval $[0,T_{max}]$ for some maximum maturity $T_{max}$.
 such that $\sigma_i$ is the volatility corresponding to the time period $(T_{i-1},T_i]$.
For simplicity, this partition corresponds to the maturities of available quoted options, hence
$N_M$ is the number of different maturities.

The calibration matches $\sigma_i$ to European At-The-Money (ATM) options' volatilities assuming constant rates\footnote{Such a step-by-step calibration is industry practice in CCR modeling.}. Denote $\tilde\sigma^{T_i}$ the implied volatility of an ATM option with maturity $T_i$, 
then the $\sigma_i$ 
can be determined inductively from the system of equations
\begin{eqnarray*}
    (\tilde\sigma^{T_i})^2T_i = \sum_{j=1}^i \sigma_j^2(T_j-T_{j-1}), \qquad i = 1,\ldots,N_M.
\end{eqnarray*}
The resulting different volatility levels are shown in Figure \ref{subfig: Vols}.

The Hull-White parameters are calibrated to fit the prevailing yield curve and swaption data using analytic expressions, i.e.\
$\Theta$ is calibrated by (\ref{eq:theta}) and $\eta$ and $\lambda$ by a least-squares fit to co-terminal swaptions that terminate in 10 years (1$\times$9, 2$\times$8, 3$\times$7, 4$\times$6, 5$\times$5, 6$\times$4, 7$\times$3, 8$\times$2 and 9$\times$1), as these swaptions can also be used to replicate the CVA of a swap that matures in 10 years, as is shown in~\citet{Sorensen1994}.

The correlations between the observable factors (exchange and interest rates) are estimated using weekly historical time-series data
from the preceding three year period.
The estimated correlation matrix is regularised to be positive semi-definite by setting any negative eigenvalues in a singular value decomposition to zero \citep{Rebonato1999}. The resulting full correlation matrix is shown in Appendix~\ref{app: Parameters} together with the other parameters.
The yield curve data are provided in a spread sheet at \cite{DimRedArXiv}.

\section{Results}\label{sec: Results}

In this section, we analyse the accuracy of the numerical approximations from Section \ref{sec: Risk factor decompostion},
applied to the market described in Section \ref{sec: Case study}.
We compare our results to an accurate approximation computed by a brute-force Monte Carlo method. In particular, we study the relative differences expressed in percentages,
\begin{eqnarray}
e_{L_2} =100\frac{\sqrt{\sum^{{\NT}}_{i=1}\left(V_{r,s}(t_i)-V_{\rm MC}(t_i)\right)^2}}{\sqrt{\sum^{{\NT}}_{i=1}V_{\rm MC}(t_i)^2}},\label{eq: eL2}
\qquad
e_{L_\infty} =100\frac{\max_{i=1}^{{\NT}} \left(\left|V_{r,s}(t_i)-V_{\rm MC}(t_i)\right|\right)}{\max_{i=1}^{{\NT}}\left(\left|V_{\rm MC}(t_i)\right|\right)},
\end{eqnarray}
where 
$r=1$ and $s=1$ or $s=2$ 
to denote two- or three-dimensional corrections (see (\ref{truncated_rs})).
For the computations, $t_i$ are chosen to coincide with the swap payment dates, so that ${\NT} = {\NC}$ is the total number of coupon payments.
In addition, we study the mean difference (MD) over time relative to the sum of all notionals ($N_{\text{\scriptsize total}}$) in the portfolio expressed in basis points,
\begin{eqnarray*}
{\rm MD} =\frac{10000}{{\NT} N_{\text{\scriptsize total}}}\sum^{{\NT}}_{i=1}\left|V_{r,s}(t_i) -V_{\rm MC}(t_i)\right|.
\end{eqnarray*}

Similarly, the normalised standard error for EE is defined as
\begin{eqnarray}
SE = 100  {\sqrt{\sum^{{\NT}}_{i=1} SE(t_i)^ 2}} \Bigg \slash{\sqrt{\sum^{{\NT}}_{i=1} EE(t_i)^2}}, \qquad
SE(t_i) = \frac{Std(EE(t_i))}{\sqrt{\Nom}},
\label{eqn:SE}
\end{eqnarray}
where $Std(EE(t_i))$ is the standard deviation of the estimator for $EE(t_i)$, and an analogous definition for EPE.

The settings used for the numerical methods (domain and mesh sizes, time steps etc) are reported in Appendix \ref{app:numparams}.

In the following, assuming piecewise volatility, we use up to seven risk factors, $d=7$, where (see (\ref{Xproc}) and (\ref{FRprocessesBS}))
\begin{eqnarray}
\label{XvsFR}
\left(X_t^1,\ldots, X_t^7\right) =
\left(
F_t^1, R^{\rm d}, R_t^{{\rm f},1}, F_t^2, R_t^{{\rm f},2}, F_t^3, R_t^{{\rm f},3}
\right)
\end{eqnarray}
are the EURUSD, EURGBP, EURJPY exchange rates ($F^i$), and EUR, USD, GBP, JPY short rates ($R^{\rm d}$ and $R^{{\rm f},i}$, respectively).

To simplify the notation (\ref{integral-v}), we will use, e.g., the shorthand
\[
V(R^{\rm d}, F^2, R^{{\rm f},3}) \equiv 
V_{\{2,4,7\}}(R^{\rm d}_0, F^2_0, R^{{\rm f},3}_0; R^{\rm d}_0, F^2_0, R^{{\rm f},3}_0,t) = V_{\{2,4,7\}}(X_0^2, X_0^4, X_0^7;X^2_0, X^4_0, X^7_0,t),
\]
where we suppress the dependence on the anchor $a=X_0$ and time $t$,
and it is understood implicitly that the arguments identify the function being used. Also note that the function is evaluated at $X_0$, since we are interested in exposures conditional on the current state $\mathcal{F}_0$.

\subsection{Case A: Single EURUSD CCYS}
\label{subsec:caseA}
As a first test case, we focus on the BS2HW model (Section \ref{subsec:drivers}) for a single CCY swap trade on EURUSD.
In this case, only the three relevant risk factors $F^1$, $R^{\rm d}$ and $R^{{\rm f},1}$ (the EURUSD exchange rate, the EUR and USD short rates) are used, so effectively $d=3$.

In that case
\begin{eqnarray}
\label{eq: BS 2D approx}
  V_{1,1} &=& V(F^1) + \left( V(F^{1},R^{\rm d}) - V(F^1) \right) + \left( V(F^{1},R^{{\rm f},1})-V(F^{1}) \right) \\
  V_{1,2}  &=&  V_{1,1} +
 \left( V(F^{1},R^{\rm d},R^{{\rm f},1}) - V(F^{1},R^{\rm d}) - V(F^{1},R^{{\rm f},1})+ V(F^1) \right) 
 \label{eq: BS 3D approx} \\
 &=& V(F^{1},R^{\rm d},R^{{\rm f},1}).
 \label{eq: BS 3D exact}
\end{eqnarray}
The first term in (\ref{eq: BS 2D approx}) is a one-dimensional approximation with only the EURUSD rate stochastic, and then the next two brackets correct separately for stochastic domestic  (EUR) and stochastic foreign (USD)  interest rate, respectively.
Note how in this particular case equation (\ref{eq: BS 3D approx}) with the three-dimensional corrections simplifies to the exact solution (\ref{eq: BS 3D exact}).

\begin{figure}[p!]
\centering
\subfigure[ EE, Case A. ]{
\includegraphics[width=0.48\textwidth]{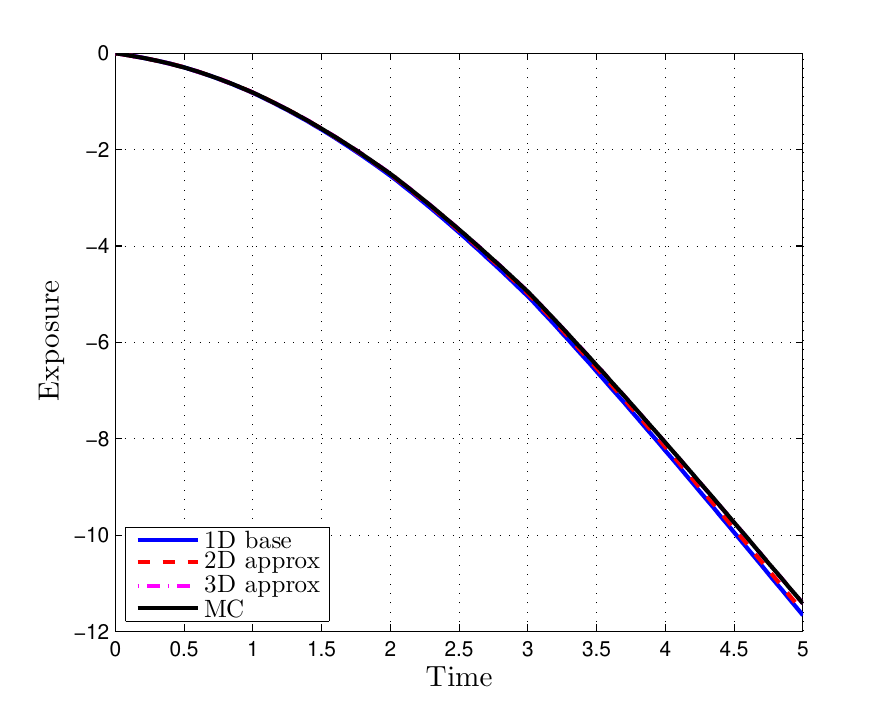}\label{subfig: BS EE single trade 2D EURUSD}
}
\subfigure[ EPE, Case A. ]{
\includegraphics[width=0.48\textwidth]{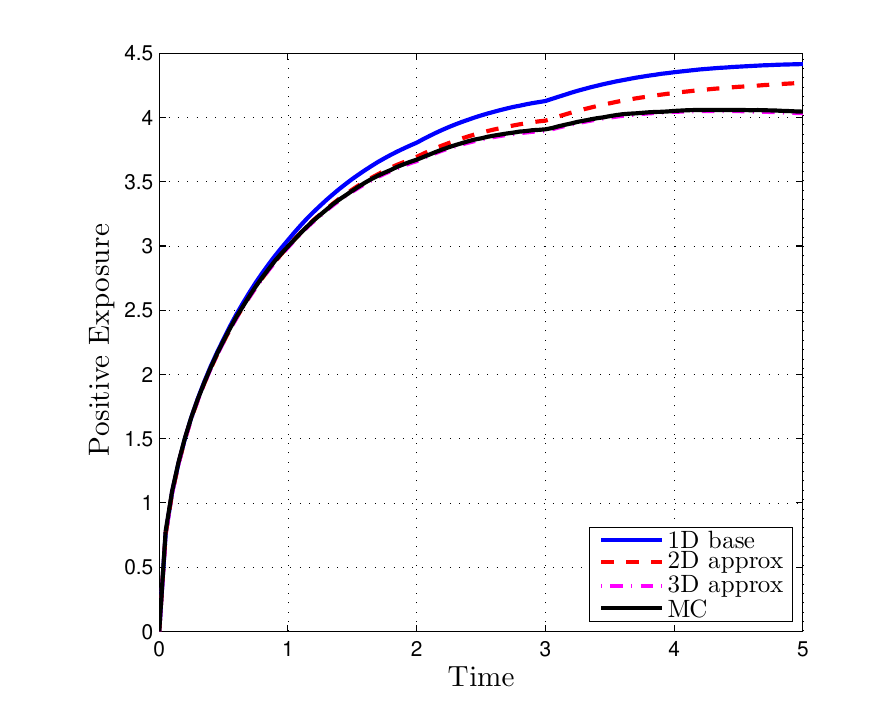}\label{subfig: BS EPE single trade 2D EURUSD}
}
\subfigure[ EE, Case B. ]{
\includegraphics[width=0.48\textwidth]{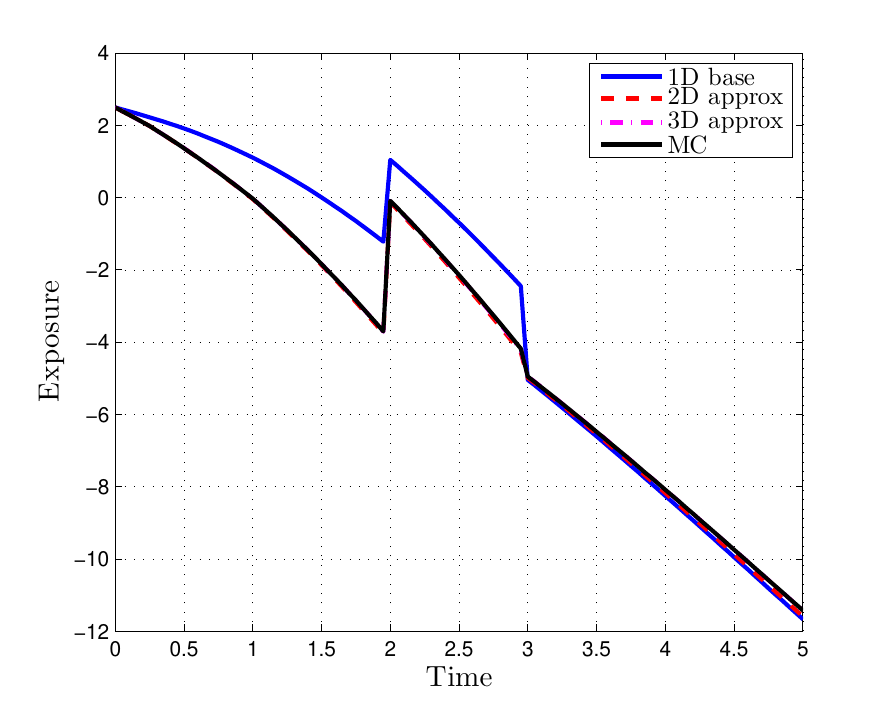}\label{subfig: BS EE port trade 2D EURUSD}
}
\subfigure[ EPE, Case B. ]{
\includegraphics[width=0.48\textwidth]{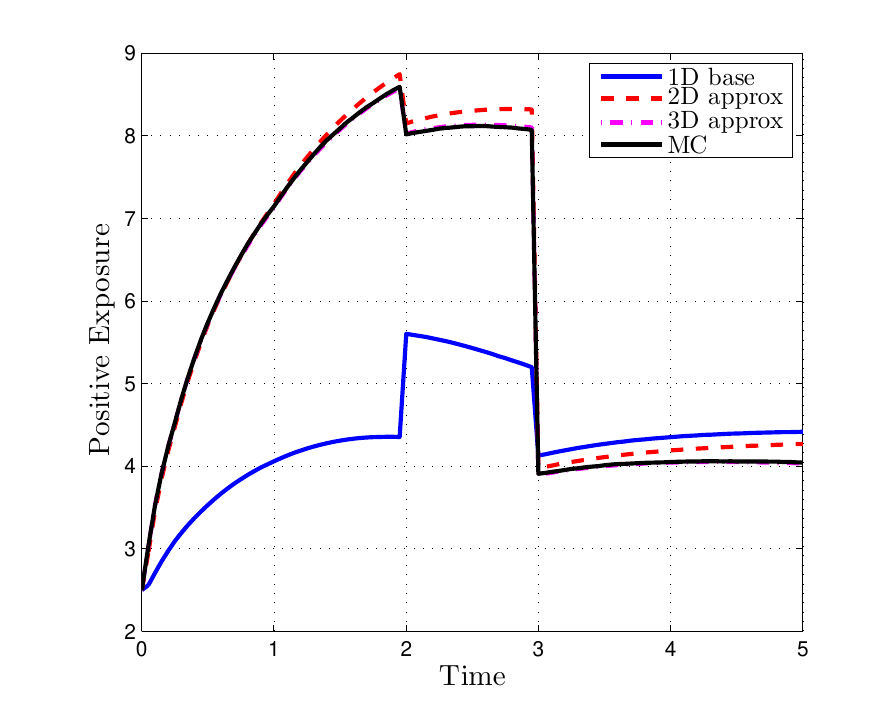}\label{subfig: BS EPE port trade 2D EURUSD}
}
\subfigure[ EE, Case C. ]{
\includegraphics[width=0.48\textwidth]{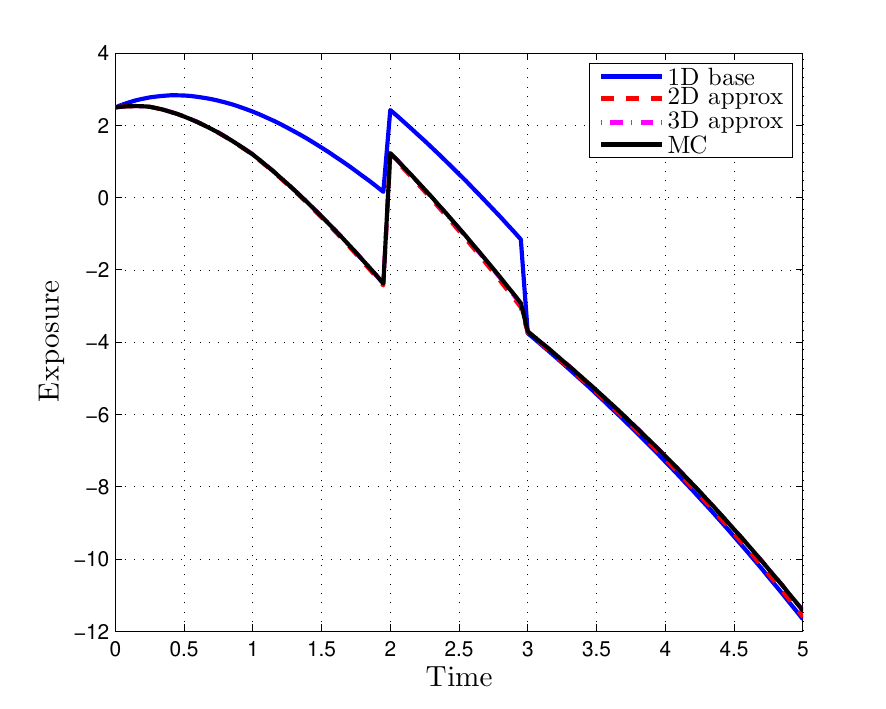}\label{subfig: EE port trade 2D EURUSD IRS}
}
\subfigure[ EPE, Case C. ]{
\includegraphics[width=0.48\textwidth]{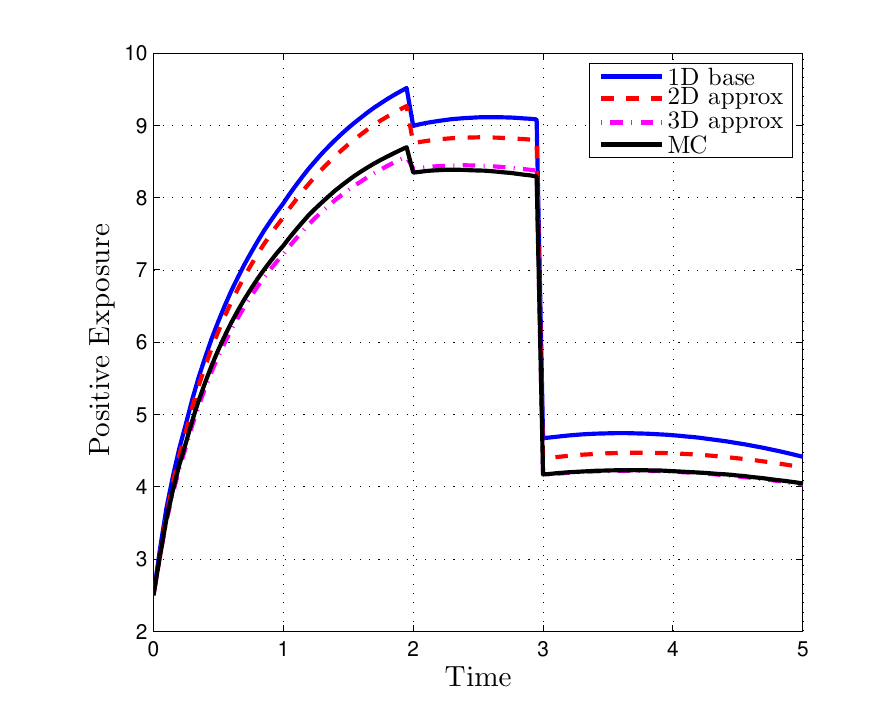}\label{subfig: EPE port trade 2D EURUSD IRS}
}
\caption{EE (left) and EPE (right) for a single ATM CCYS (\ref{subfig: BS EE single trade 2D EURUSD} and \ref{subfig: BS EPE single trade 2D EURUSD}), for three CCYS (\ref{subfig: BS EE port trade 2D EURUSD} and \ref{subfig: BS EPE port trade 2D EURUSD}), and for three CCYS and an IR swap (\ref{subfig: EE port trade 2D EURUSD IRS} and \ref{subfig: EPE port trade 2D EURUSD IRS}). The risk factors are driven by BS2HW and the EURUSD FX rate is taken as a base.  }
\end{figure}

In Figures \ref{subfig: BS EE single trade 2D EURUSD} and \ref{subfig: BS EPE single trade 2D EURUSD}, the one-,
 two- and three-dimensional decomposed approximations $V_{1,0}$, $V_{1,1}$ and $V_{1,2}$
are plotted together with the full scale Monte Carlo approximation $V_{\rm MC}$. The exposure decreases over time,
which means that the FX rate drives the expected value of this swap down. The EPE is positive by definition and increases over time.

Both the decomposed two- and three-dimensional approximations are reasonably close to the full Monte Carlo estimator, but  while the results for Case A in Table \ref{Table: error table} show that the maximum error is over 5\% in the two-dimensional case, it is around 0.5\% for the three-dimensional approximation.
Note that the latter result should be exact, so that any discrepancy are finite difference errors. These may seem large, but we note that the standard error of Monte Carlo with the industry standard of $10^4$ paths would be approximately $0.1\% \sqrt{4\cdot10^6/10^4} = 2\%$ (see Table \ref{Table: error table}). While we could certainly improve the finite difference accuracy by finer meshes, we do not investigate this further at this point.\footnote{See Appendix \ref{app: FD Errors} for an analysis of the accuracy of the finite difference schemes used.}

\begin{table}[ht!]
\begin{center}
\caption{Errors of exposure approximations for Cases A, B and C. The finite difference approximations are computed with $m_1 = 60$ grid points and 500 time steps. The errors are expressed in percentages together with the standard error of the Monte Carlo benchmark with $4 \cdot 10^6$ paths and 1000 time steps. The SE is defined in (\ref{eqn:SE}) as the root sum squared of the standard errors over time relative to the root sum squared of the sampled EE or EPE.\label{Table: error table}}
{
{\begin{tabular}{{llccc|ccc|ccc|c}}\toprule
            & &  \multicolumn{3}{c}{$e_{L_2}$ (\%)}& \multicolumn{3}{c}{$e_{L_\infty}$ (\%)}& \multicolumn{3}{c}{MD (bp)}& SE (\%)\\ \toprule
            	& 		      & 1D    & 2D  & 3D    & 1D   &   2D & 3D   & 1D & 2D & 3D & \\ \hline
\multirow{2}{*}{Case A} & EE        & 2.07  & 1.20&0.038 & 2.16 & 1.50 &0.049 & 8.58& 4.43 & 0.17 & 0.13    \\
                        & EPE       &  6.11 & 2.83& 0.14 & 9.27 & 5.66 & 0.55 &18.29& 7.35 & 0.40 & 0.085\\ \midrule
\multirow{2}{*}{Case B} & EE        & 19.54 & 1.41& 0.20 & 21.67 & 1.50 & 0.34 & 32.46 & 2.52 & 0.25 & 0.17    \\
                        & EPE       & 36.51 & 2.29& 0.26 & 49.44 & 3.09 & 0.54 & 359.1& 4.93 & 0.46 & 0.075 \\ \midrule
\multirow{2}{*}{Case C} & EE        & 22.23  & 1.48& 0.18&  22.19 & 1.44 & 0.30 &  33.26 & 2.36& 0.22 & 0.19   \\
                        & EPE       & 29.83  & 5.63& 1.17&  41.87 & 6.85 & 1.57 &  324.7 & 13.24& 2.32 & 0.073  \\ \bottomrule 
\end{tabular}}
}
\end{center}
\end{table}


\subsection{Case B: Three CCYS (EURUSD, EURGBP and EURJPY)}
\label{subsec:caseB}
In the case of three CCYS, there are $d=7$ risk factors, such that adding three-dimensional corrections will no longer yield an exact solution. In this case, we make an \emph{a priori} choice of the base risk factor, namely the EURUSD FX rate.\footnote{See Section \ref{subsec:otherbases} for an assessment of alternative choices.} The two-dimensional decomposed approximation for the portfolio of three FX swaps now has two extra terms to correct for the other FX rates and two extra terms to correct for the foreign interest rates:
\begin{eqnarray}
\nonumber
  V_{1,1} =& V(F^{1})&+\left( V(F^{1}, F^2)-V(F^{1})\right)+\left( V(F^{1},F^3)-V(F^{1}) \right) \nonumber + \left( V(F^{1},R^{\rm d})-V(F^{1})\right) 
  \\
 & &+\left( V(F^{1},R^{{\rm f},1})-V(F^{1})\right) + \left( V(F^{1},R^{{\rm f},2})-V(F^{1})\right) + \left( V(F^{1},R^{{\rm f},3})-V(F^{1})\right)\nonumber \\
 = &V(F^{1})&+ \sum_{j=2,3} \Delta V(F^{1}, F^{j}) + \Delta V(F^{1}, R^{\rm d}) +  \sum_{j=1,2,3} \Delta V(F^{1},R^{{\rm f},j}), \label{eq: 2D corrections}
\end{eqnarray}
where the short-hand $\Delta V(X^{1}, X^{j}) \equiv V(X^{1}, X^{j}) - V(X^{1})$ was used in the last line.

For the three-dimensional corrections, there are in principle $\binom{6}{2}=15$ extra terms, because in addition to $X_1$, we choose 2 out of the $d-1=6$ remaining factors.
However, only 8 of them are non-zero, as given here:
%
 \begin{eqnarray}
 \label{eqn:V12}
  V_{1,2}= &V_{1,1}&+\Delta V(F^{1},F^{2},F^{3}) +\Delta  V(F^{1},F^{2},R^{{\rm d}}) +\Delta V(F^{1},F^{3},R^{{\rm d}}) \\
    & &+\sum_{j=1,2} \Delta V(F^{1},F^{2},R^{{\rm f},j}) 
    +\sum_{j=1,3} \Delta V(F^{1},F^{3},R^{{\rm f},j}) 
    + \Delta V(F^{1},R^{{\rm d}},R^{{\rm f},1}), \nonumber 
 \end{eqnarray}
 where the short-hand $\Delta V(X^{1}, X^{i}, X^{j}) \equiv V(X^{1}, X^{i}, X^{j}) -V(X^{1}, X^{i})  - V(X^{1}, X^{j}) + V(X^{1})$ was used.


The corrections can be interpreted in the following way.
For instance, in the first term in (\ref{eqn:V12}), $V(F^{1},F^{2},F^{3})$ treats all exchange rates stochastic but with deterministic interest rates (simple Black-Scholes models); and in the last term, $V(F^{1},R^{{\rm d}},R^{{\rm f},1})$ comes from the full BS2HW model for the EURUSD rate where all other processes are approximated by their expectations (see Section \ref{subsec-anova}). Similar interpretations can be found for the other correction terms.
 
In contrast, we have for instance
\[
\Delta  V(F^{1},R^{{\rm d}},R^{{\rm f},3}) = 
\underbrace{V(F^{1}, R^{{\rm d}},R^{{\rm f},3}) -V(F^{1}, R^{{\rm d}}) }_{=0}
\underbrace{- V(F^{1}, R^{{\rm f},3}) + V(F^{1})}_{=0} = 0,
\]
because in this approximation where (the EURJPY exchange rate) $F^{3}$ is deterministic, the (JPY) short rate $R^{{\rm f},3}$ has no impact on the exposure of the swaps. Similar arguments hold for the other 6 corrections that vanish.

Figures \ref{subfig: BS EE port trade 2D EURUSD} and \ref{subfig: BS EPE port trade 2D EURUSD} show again the one- to three-dimensional decomposed approximations of the exposures together with the full scale Monte Carlo approximation $V_{\rm MC}$. The different expiries for the different swaps are reflected clearly in the profiles. At $T=2$, the EURJPY swap expires, which results in an upwards jump in total EE and EPE whereas at $T=3$ the EURGBP swap expires and the exposures drop. 

Both the two- and especially the three-dimensional approximation are close to the full Monte Carlo estimator; see especially the rows relating to Case B in Table \ref{Table: error table}. 
The mean differences are now relative to the sum of all notionals in the portfolio.
As none of the correction terms accounts simultaneously for, say, the EURJPY rate, EUR and JPY short rates, even the decomposition with three-dimensional approximations is not exact in this case, such that all errors arise from a combination of truncation error for the expansion and discretisation error for the PDE solution (as well as a slightly smaller Monte Carlo error for the benchmark).

We give a table of all the correction terms separately in Appendix \ref{sec:corrections}. Table \ref{Table: corrections FDs EE} there shows that while some terms are dominant, like the one involving the EURUSD and EURGBP rates among the two-dimensional corrections and the one with EURUSD and the associated short rates among the three-dimensional ones, it is certainly not the case that the other terms are negligible. This demonstrates that the problem is genuinely high-dimensional and does not have a lower superposition dimension (see \citet{wang2005high}), by which we mean loosely speaking that the solution cannot be expressed exactly as a linear combination of solutions to low-dimensional problems. Indeed, while the approximations $V_{0,3}$ and hence $V_{1,3}$ would be exact for EE here because of its linearity, $V_{1,2}$ is not. For EPE, generally only
decompositions involving all risk factors, such as $V_{0,7}$ or any $V_{r,s}$ with $r+s=7$ will be exact. The results show, however, that approximations with much smaller $r$ and $s$ and therefore much lower computational cost can be sufficiently accurate.

In Section \ref{subsec:varred} we will show the variance reduction achieved in Case B by using the above approximation as control variate.


\subsection{ Case C: Case B with an additional IRS in EUR}
\label{subsec:caseC}
By adding an IR swap to the portfolio, the risk factors do not change, as the EUR interest rate was already modeled as a factor that drives the FX rate in Section \ref{subsec:caseB}. In Figures \ref{subfig: EE port trade 2D EURUSD IRS} and \ref{subfig: EPE port trade 2D EURUSD IRS}, one observes the exposure levels are increased because of the interest rate swap. Again, the different profiles match closely over time. In Table \ref{Table: error table} one sees more clearly that, for all error measures, both the EE and EPE errors improve significantly when we include three-dimensional corrections, as expected.

One thing to note is that the errors for EE are almost exactly the same as in Case B, because the IR swap exposure is valued exactly by all models that have $R^{\rm d}$ in them, e.g., $V(F^1,R^{\rm d})$.

\subsection{Case D: Case C with an additional EURUSD FX call option}

\begin{figure}[ht!]
\centering
\subfigure[ EE, Case D. ]{
\includegraphics[width=0.46\textwidth]{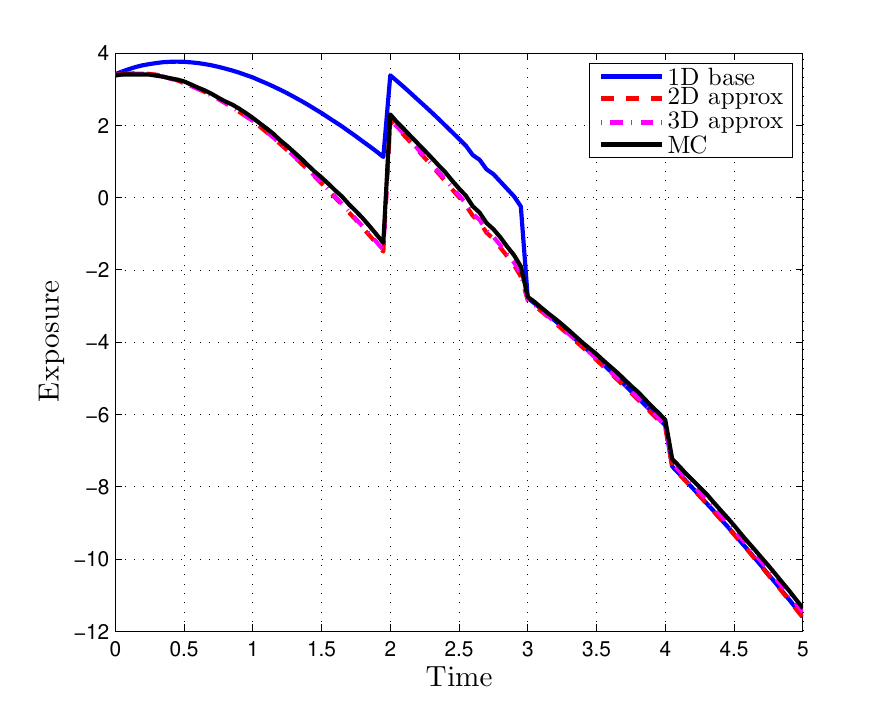}\label{subfig: EE port trade 2D EURUSD Call IR}
}
\subfigure[ EPE, Case D. ]{
\includegraphics[width=0.46\textwidth]{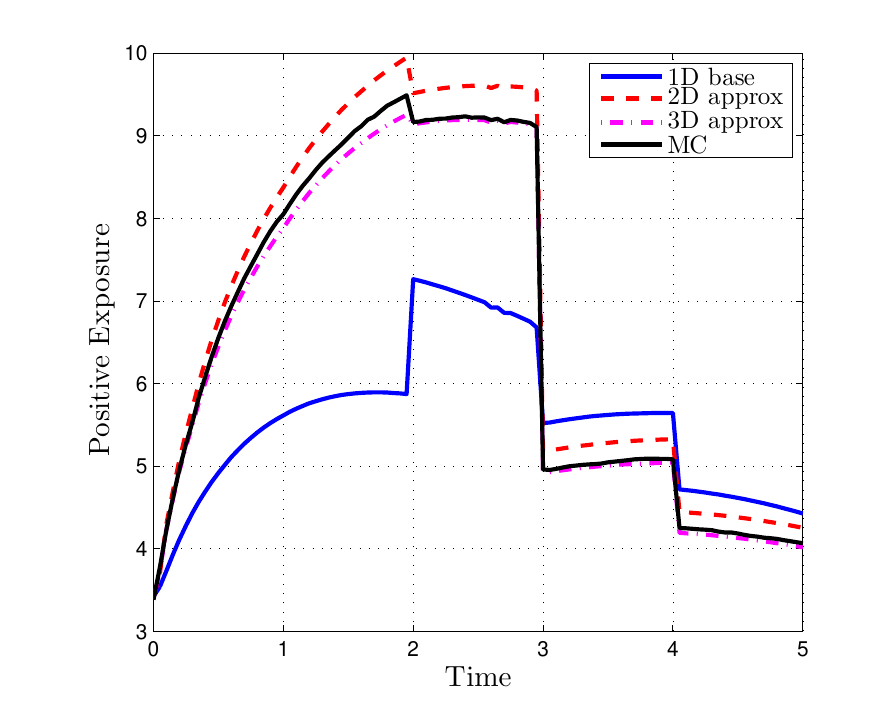}\label{subfig: EPE port trade 2D EURUSD Call IR}
}
\caption{EE (left) and EPE (right) for a portfolio of CCYS, a call option on the EUR USD FX rate and a floating vs fixed interest rate swap. The factors are driven by BS2HW and the EURUSD FX rate is taken as the base factor.  }\label{fig: port trade PCAs Call IR}
 \end{figure}

We now add an FX option on the EURUSD rate as described in Section \ref{subsec: Derivatives}.
The exposure level increases first, but drops at $t = 4$, as seen in Figures \ref{subfig: EE port trade 2D EURUSD Call IR} and \ref{subfig: EPE port trade 2D EURUSD Call IR}. 

In the previous examples, Sections \ref{subsec:caseA}--\ref{subsec:caseC}, analytical pricing formulas were available for the swaps in the portfolio, which made Monte Carlo estimation of exposures straightforward. With the FX option added, we use the regression-based MC method, similar to the popular Longstaff-Schwartz algorithm \citep{Longstaff2001} for American options.
An outline of the algorithm used is given in Appendix \ref{sec:regression}.

Because of this regression, all paths need to be stored and less paths can be used, which results in a higher standard error. 
We also note that the benchmark in this case is not perfect due to the limited span of the basis functions. The resulting bias is not taken into account in the SE given in Table \ref{Table: error port trade calls IRS}.
Nonetheless, the discrepancy between PDE results and Monte Carlo behaves as expected -- see Table \ref{Table: error port trade calls IRS} -- with a sharp improvement as corrections are added in. 
\begin{table}[ht!]
\begin{center}
\caption{Errors of exposures of a portfolio with three CCYS, a IRS and a FX call option. The errors are expressed in percentages together with the standard error of the Monte Carlo benchmark with $2\cdot 10^5$ paths. The SE is defined in (\ref{eqn:SE}) as the root sum squared of the standard errors over time relative to the root sum squared of the sampled EE or EPE.\label{Table: error table FX call}}
{\begin{tabular}{@{}lccc|ccc|ccc|c}\toprule
            &   \multicolumn{3}{c}{$e_{L_2}$ (\%)}&  \multicolumn{3}{c}{$e_{L_\infty}$ (\%) }&  \multicolumn{3}{c}{MD (bp)} & SE (\%)\\ \toprule
            &  1D   &  2D  & 3D    & 1D    & 2D   & 3D    & 1D   &  2D  & 3D  & MC    \\ \midrule	
EE          & 23.20\  & 1.43   & 0.32   & 22.57   & 1.56   & 0.36   & 33.72  & 2.08   & 0.52 & 0.86\\
EPE         & 26.36\  & 4.83   & 1.29   & 37.87   & 5.44   & 1.87   & 319.00 & 12.51  & 2.62 & 0.30\\\bottomrule
\end{tabular}}
 \label{Table: error port trade calls IRS}
\end{center}
\end{table}




\subsection{Stochastic volatility}
\label{subsec:Stochastic Volatility}
To investigate the applicability of this method to stochastic volatility, we model all three FX rates with the Heston model \citep{Heston1993}, where the stochastic volatility is modeled by a CIR process:
\begin{subequations}
\label{FRprocessesHes}
    \begin{eqnarray}
         dF^i_t   =& (R_t^{\text{d}}-R_t^{\text{f},i}) F^i_t \, dt + \sqrt{Y^i_t} F^i_t \, dW^{{\rm F},i}_t, \,\;\qquad\qquad\qquad \qquad \qquad \qquad 1\le i \le m, \label{eq:SDE SV}\\
         dY^i_t =& \kappa_i\left(\bar{v}_i-Y^i_t\right)dt+\gamma_i \sqrt{Y^i_t}dW^{{\rm Y},i}_t, \,\; \quad\qquad\qquad\qquad \qquad \qquad \qquad 1\le i \le m, \label{eq:SDE SV Y} \\
         dR^{\text{d}}_t   =& \lambda_{\text{d}}(\Theta_{\text{d}}(t)-{R}^{\text{d}}_t  ) \, dt + \eta_{\text{d}} \, dW^{\rm d}_t, \quad \qquad \qquad \qquad \qquad\qquad\qquad \qquad \qquad \qquad
         \label{eq:SDE r_d SV}\\
         dR^{\text{f},i}_t   =& \left[\lambda_{\text{f}}^i (\Theta_{\text{f}}^i(t)-R^{\text{f},i}_t  )-\eta^i_{\text{f}} \rho_{I(i), J(i)} 
         \sqrt{Y^i_t} \right] \, dt + \eta_{\text{f}}^i \, dW^{{\rm f},i}_t,  
         \quad \qquad \qquad \quad 1\le i\le m, \label{eq:SDE r_f SV}    
         \end{eqnarray}
\end{subequations}
with $Y^i_0 =v_{0,i}$ given, and other notation as earlier.
The three sets of Heston parameters are calibrated to volatility smiles from 2 December 2014 and are given with the calibration fit in Appendix \ref{app: Parameters Heston}.  
It is unclear how to estimate the correlations between the stochastic variance and the other risk factors (apart from its own FX rate).
In this test, they are assumed to be zero.\footnote{After correcting the estimated full correlation matrix for positive-definiteness, these correlations can be slightly non-zero. The full correlation matrix used in this test can be found in equation (\ref{eq: Heston correlations}) in Appendix \ref{app: Parameters Heston}.}
The extension to local-stochastic volatility models would not cause any computational difficulty and the accuracy of the approximation is expected to be similar.

In this case, there are $d=10$ risk factors and we again choose the EURUSD FX rate \emph{a priori} as base risk factor. We consider Case B with three  CCYS with different moneyness and maturities.

The two-dimensional decomposed approximation for the portfolio of three FX swaps now has one extra term to correct for the stochastic volatility of the EURUSD base risk factor:
\begin{eqnarray}
\nonumber
  V_{1,1} 
 = &V(F^{1})&+ \sum_{j=2,3} \Delta V(F^{1}, F^{j}) + \Delta V(F^{1}, R^{\rm d}) +  \sum_{j=1,2,3} \Delta V(F^{1},R^{{\rm f},j})+\Delta V(F^{1}, Y^{1}) , \nonumber
\end{eqnarray}
where the short-hand notation from equation (\ref{eq: 2D corrections}) was used. 

For the three-dimensional corrections, in this case there are in principle $\binom{9}{2}=36$ extra terms, because in addition to $X^1$, we choose 2 out of the $d-1=9$ remaining factors. In this preliminary test, however, we only add a stochastic volatility factor to the four 2D corrections that contribute the most in the Black-Scholes case.\footnote{See Table \ref{Table: corrections FDs EE} in Appendix \ref{sec:corrections}.} These include corrections for the volatility in the two additional FX rates (EURGBP and EURJPY):
 \begin{eqnarray}
 \label{eqn:V12 hes}
  V_{1,2}= &V_{1,1}&+\Delta V(F^{1},F^{2},F^{3}) +\Delta  V(F^{1},F^{2},R^{{\rm d}}) +\Delta V(F^{1},F^{3},R^{{\rm d}}) \\
    & &+\sum_{j=1,2} \Delta V(F^{1},F^{2},R^{{\rm f},j}) 
    +\sum_{j=1,3} \Delta V(F^{1},F^{3},R^{{\rm f},j}) 
    + \Delta V(F^{1},R^{{\rm d}},R^{{\rm f},1}), \nonumber \\
    & &+\sum_{j=1,2} \Delta V(F^{1},F^{j},Y^{j}) 
    +\Delta V(F^{1},Y^{1},R^{{\rm d}}) 
    + \Delta V(F^{1},Y^{1},R^{{\rm f}}), \nonumber 
 \end{eqnarray}
 where the short-hand notation from equation (\ref{eqn:V12}) was used.


Note that all the corrections without stochastic volatility treat all exchange rates stochastic but with a time dependent volatility function equal to the expectation of the future volatility,
which can be  calculated analytically as
\begin{align}\mathbb{E}\left[Y^i_t|Y^i_0\right] = v_{0,i}e^{ - \kappa_i t} + \bar{v}_i  (1 - e^{-\kappa_i t}). \label{eq: mean volatility espr}
\end{align}

\begin{figure}[h!]
\centering
\subfigure[ EE under Heston, Case B. ]{
\includegraphics[width=0.48\textwidth]{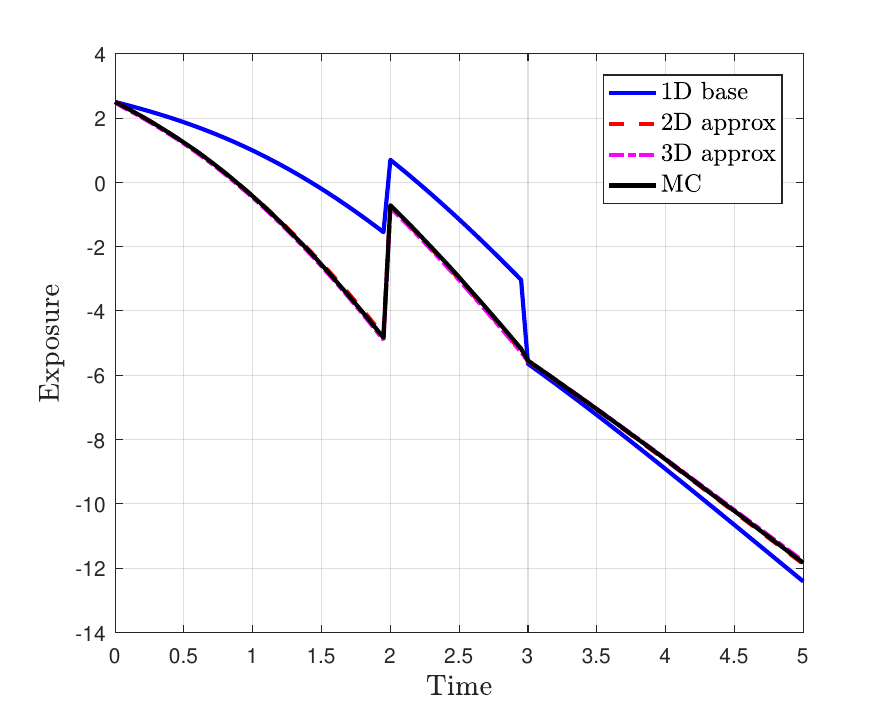}\label{subfig: Hes EE Case B EURUSD}
}
\subfigure[ EPE under Heston, Case B. ]{
\includegraphics[width=0.48\textwidth]{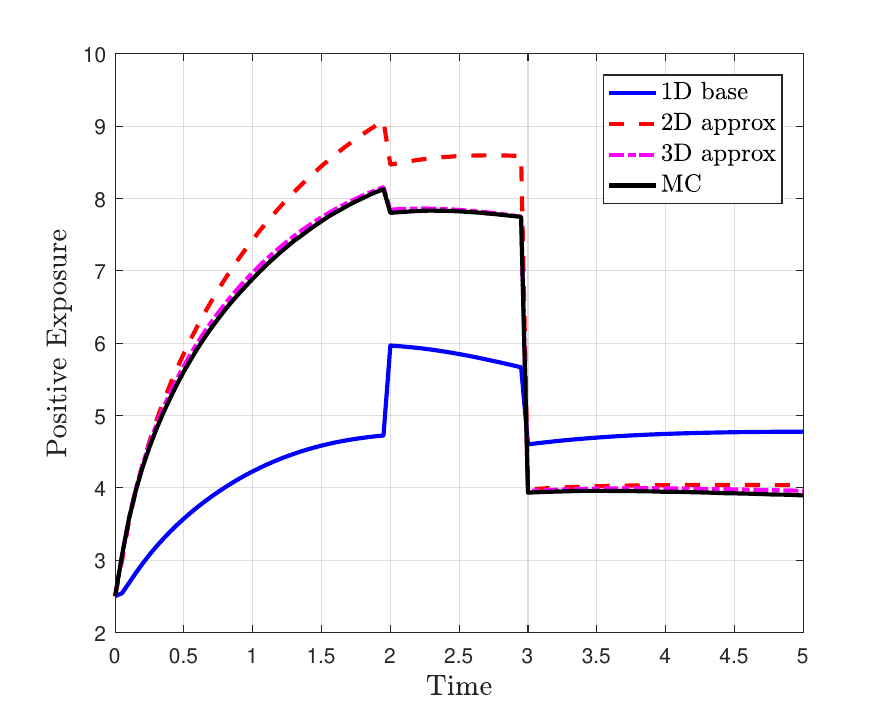}\label{subfig: Hes EPE Case B EURUSD}
}
\caption{EE (left) and EPE (right) for three CCYS under Heston.
The EURUSD FX rate is taken as a base risk factor and all FX rates are simulated with stochastic volatility.  }
\end{figure}

The one- to three-dimensional decomposed approximations of the exposures including stochastic volatility are shown in Figures \ref{subfig: Hes EE Case B EURUSD} and \ref{subfig: Hes EPE Case B EURUSD}, together with the full scale Monte Carlo approximation $V_{\rm MC}$. Again, the profiles reflect the different expiries, and we see that the 
two-dimensional approximation significantly outperforms the one-dimensional approximation, and for EPE the 
three-dimensional approximation significantly outperforms the two-dimensional approximation.

This is also clearly visible in Table \ref{Table: error table Heston}, where we show various error measures as earlier for the different approximations.
In the case of expected exposure, it appears that the two-dimensional decomposed approximation is already close to the full-scale Monte Carlo solution. This is in line with \cite{Simaitis2016}, where it is found that for expected exposure, a time dependent volatility function can be used, whereas only for non-linear exposures (such as expected positive or negative exposure), a full stochastic volatility model should be used.

\begin{table}[ht!]
\begin{center}
\caption{Errors of exposure approximations for Case B modeled with the Heston model. The finite difference approximations are computed with $m_1 = 80$ grid points and 500 time steps. The errors are expressed in percentages together with the standard error of the Monte Carlo benchmark with $4\cdot10^6$ paths and 1000 time steps. The SE is defined in (\ref{eqn:SE}) as the root sum squared of the standard errors over time relative to the root sum squared of the sampled EE or EPE.\label{Table: error table Heston}}
{\begin{tabular}{{llccc|ccc|ccc|c}}\toprule
            & &  \multicolumn{3}{c}{$e_{L_2}$ (\%)}& \multicolumn{3}{c}{$e_{L_\infty}$ (\%)}& \multicolumn{3}{c}{MD (bp)}& SE (\%)\\ \toprule
            	& 		      						& 1D    & 2D  	& 3D    & 1D   &   2D & 3D   & 1D & 2D & 3D & \\ \hline
& EE        & 23.40 & 0.62	&0.68  & 28.03 & 0.87 &0.75 &42.79& 1.09 & 1.35 & 0.18   \\
                        				& EPE       & 31.84 & 8.55	&0.96  & 41.70 & 12.00 &1.08 &368.01& 15.59 & 2.13 & 0.076\\ \midrule
\end{tabular}}
\end{center}
\end{table}

\subsection{Variance reduction}
\label{subsec:varred}

\begin{figure}[p!]
\centering
\subfigure[ Variance reduction for EE, single CCYS]{
\includegraphics[width=0.48\textwidth]{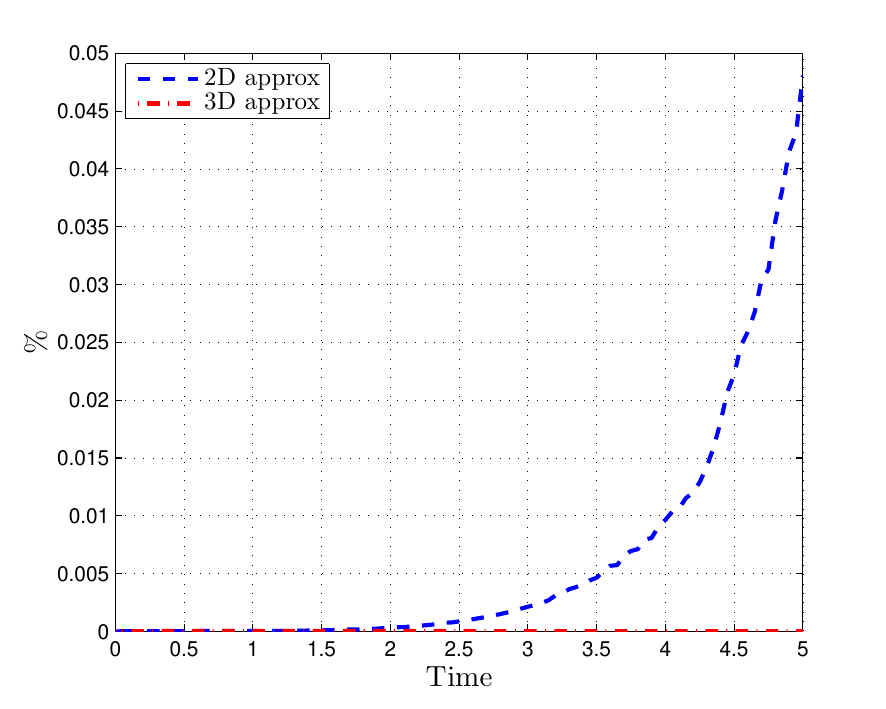}\label{subfig: var reduc EE MC A}
}
\subfigure[Variance reduction for EPE, single CCYS ]{
\includegraphics[width=0.48\textwidth]{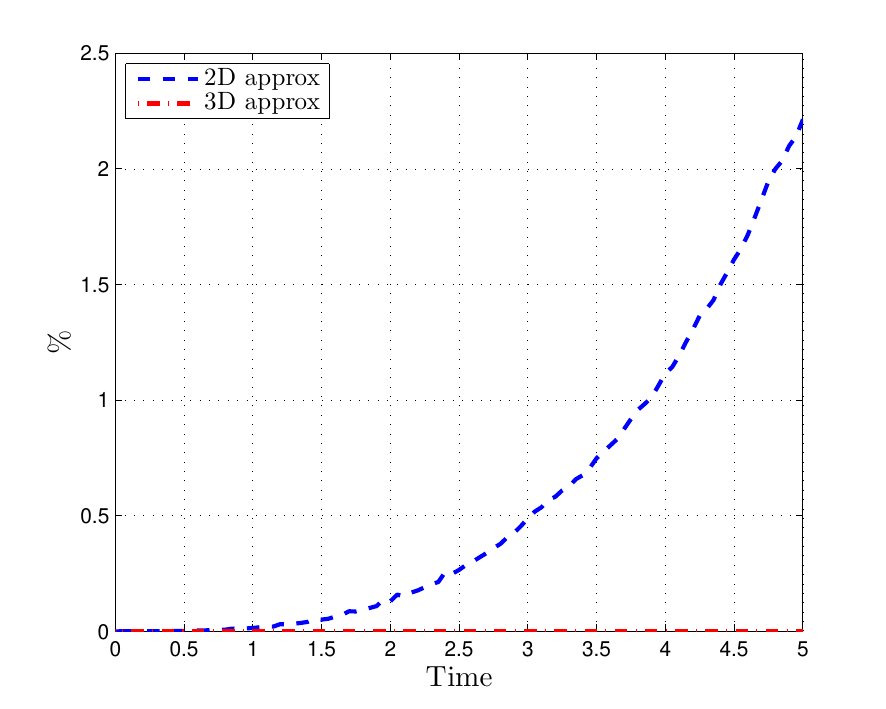}\label{subfig: var reduc EPE MC A}
}
\subfigure[Variance reduction for EE, three CCYS ]{
\includegraphics[width=0.48\textwidth]{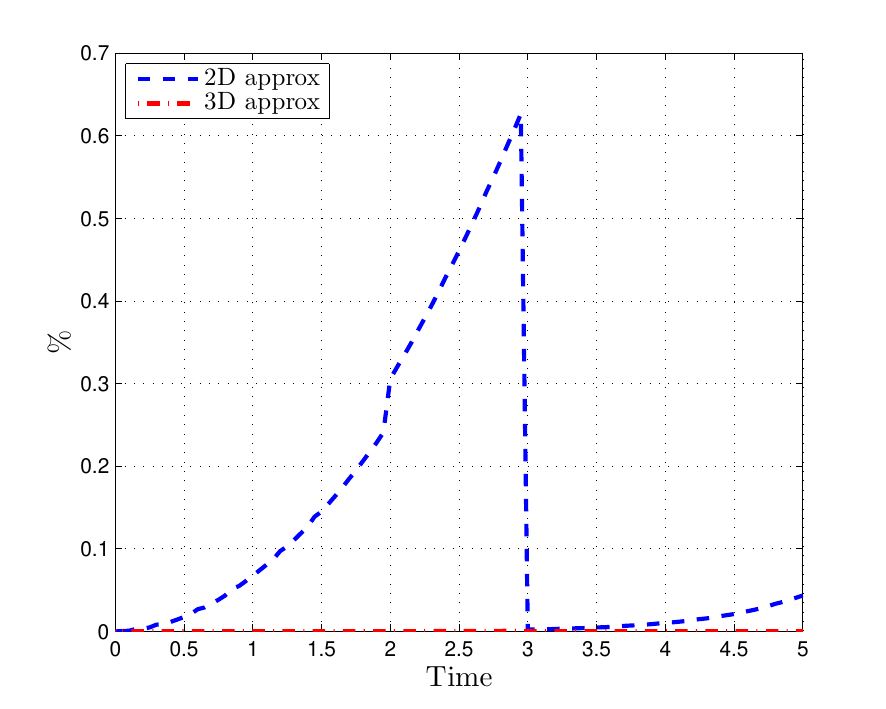}\label{subfig: var reduc EE MC B}
}
\subfigure[ Variance reduction for EPE, three CCYS ]{
\includegraphics[width=0.48\textwidth]{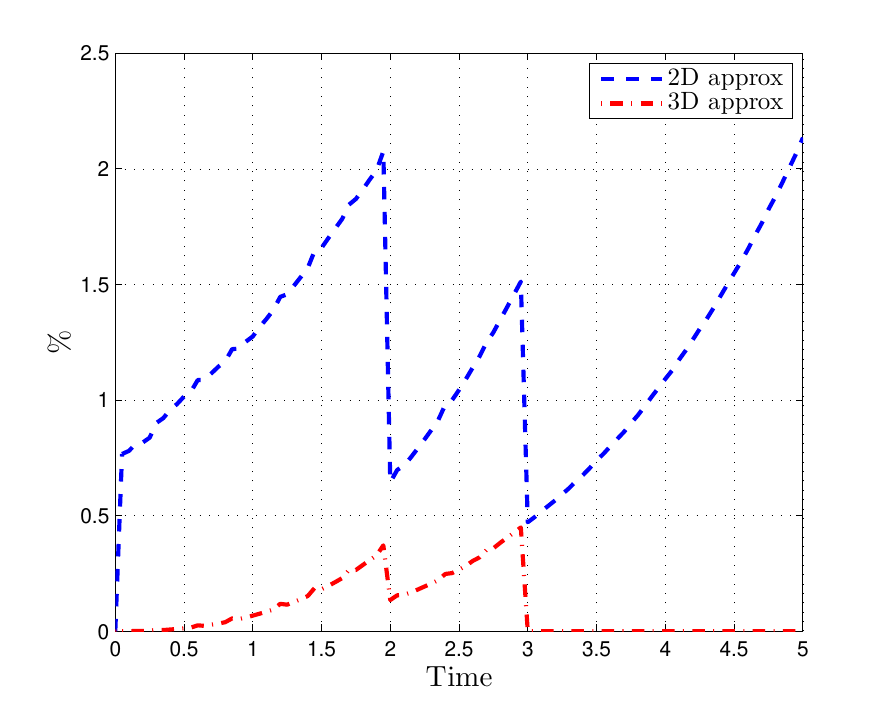}\label{subfig: var reduc EPE MC B}
}
\centering
\subfigure[ Variance reduction for EE, three CCYS and IRS ]{
\includegraphics[width=0.48\textwidth]{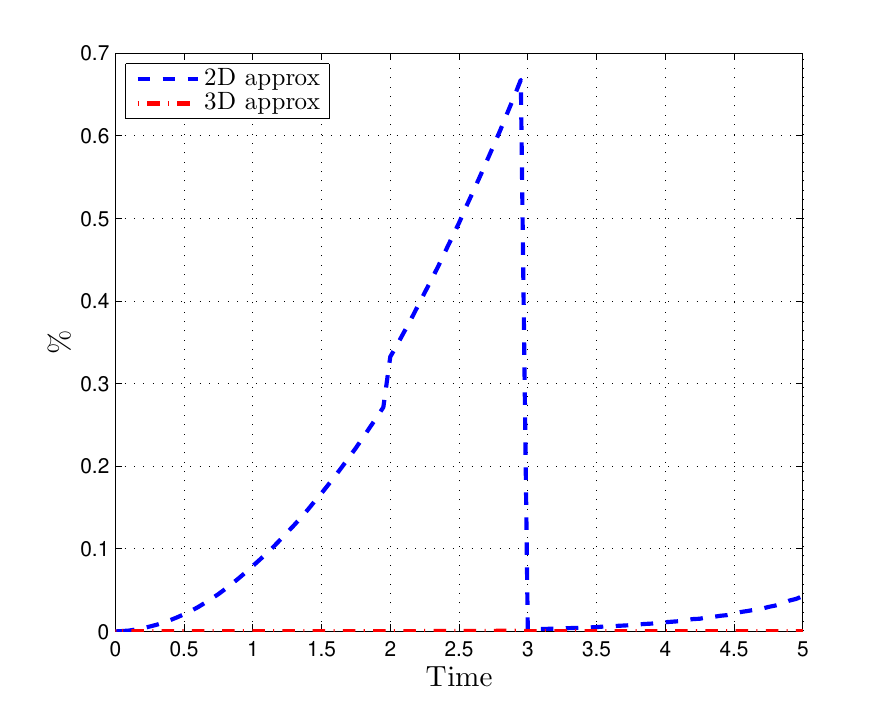}\label{subfig: var reduc EE MC C}
}
\subfigure[ Variance reduction for EPE, three CCYS and IRS ]{
\includegraphics[width=0.48\textwidth]{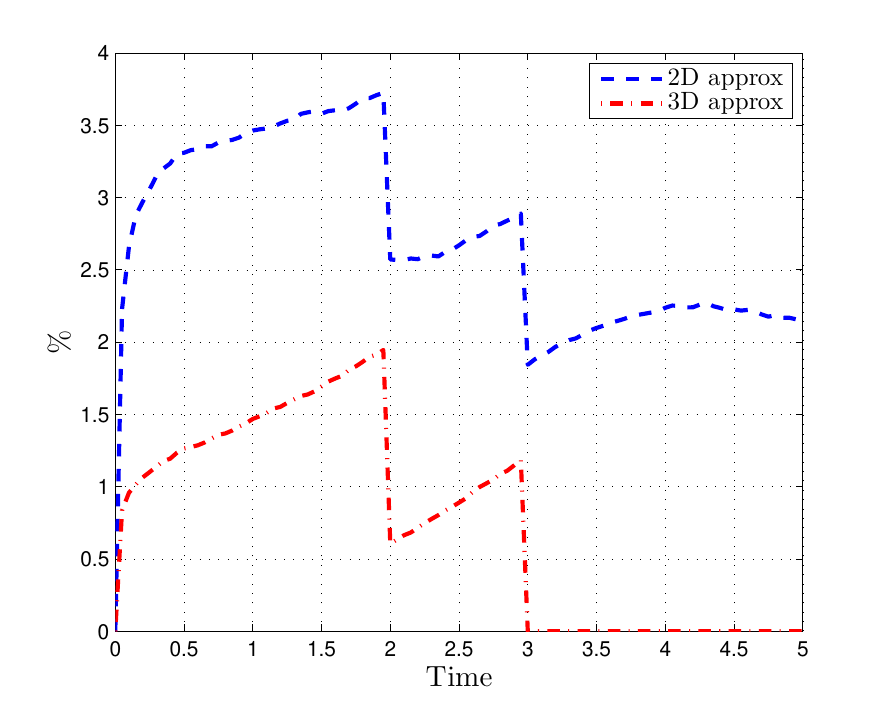}\label{subfig: var reduc EPE MC C}
}
\caption{  Variance reduction for different test cases with 2D and 3D corrections. The variance with and without control variate is compared. For this computation $10^6$ paths were used. }\label{fig: var reductions MC}
 \end{figure}

We now correct the bias in the results in the preceding sections by applying the methodology of Section \ref{subsec:cv},
at much reduced variance compared to the standard estimator.
For simplicity, we restrict ourselves to the case where derivatives are priced analytically, so that no regression is required.
The obtained variance reduction, calculated as the ratio of the two variances with and without control variate, is shown in Figure \ref{fig: var reductions MC} for the different decomposed approximations and test cases from Sections \ref{subsec:caseA} to \ref{subsec:caseC}. To be clear, a value of, e.g., $0.5\%$ means that the variance is reduced by a factor of 200, while a value of $100\%$ implies that no variance reduction at all was achieved.

In Figures \ref{subfig: var reduc EE MC A} and \ref{subfig: var reduc EPE MC A}, the variance reduction factor for EE and EPE in the case of a single CCYS is in the range of 100 to 1000. For the two-dimensional control variate, the reduction is greater for shorter maturities and increases with time. 
For the three-dimensional control variate, the variance reduction is constant, which is due to the exactness of the control variate. The only discrepancy between the control variate and the Monte Carlo estimator is the (time) discretization error, which also explains why the variance is not equal to zero.

Figures \ref{subfig: var reduc EE MC B} and \ref{subfig: var reduc EPE MC B} show the variance reduction in the case of three CCYS. Here, the three-dimensional decomposed approximation is no longer exact and thus, also in the three-dimensional case, the variance increases over time. The sharp drop at $T=2$ and $T=3$ is due to the expiry of the EURJPY and EURGBP CCYS. The variance reduction is largest between time 3 and 5 because during that time period only the EURUSD CCYS is not terminated.
When we use two-dimensional corrections, the reduction is around 200 for EE and 50 for EPE. When three-dimensional corrections are taken into account, we obtain a reduction by a factor of $10^5$ for EE and 200 for EPE.\footnote{The extreme variance reduction for EE and 3D is more clearly seen quantitatively from Figure \ref{fig: other PCAs} later on, where the data are plotted on a log scale.}

Figures \ref{subfig: var reduc EE MC C} and \ref{subfig: var reduc EPE MC C} show the variance reduction in the case of three CCYS and an IRS. The EE results resemble those for three CCYS alone, because the stochasticity of the IRS due to $R^{\rm d}$ is modelled exactly by the corrections, as explained at the end of Section \ref{subsec:caseC}.
For EPE, the variance is reduced by a factor around 30 for two-dimensional corrections and a factor of 50 for three-dimensional corrections. 

The set-up is chosen so that the finite difference accuracy is comparable to the Monte Carlo accuracy. As the finite difference approximation converges at a higher order (for low dimensions, typically up to about 3 or 4) than the Monte Carlo sampling,  the computational effort of the Monte Carlo component will become dominant if higher accuracy is required.
As a side note, we remark here that the different PDEs and integration problems can be solved by entirely different methods, e.g., Fourier methods could be used for some of the lower-dimensional problems or quasi-Monte Carlo methods for some of the medium-dimensional problems.\footnote{Similarly, quasi-Monte Carlo could be used in the benchmark method and offers a potential reduction in simulation paths from $O(n^4)$ to $O(n^2)$ for simulation-in-simulation (which is rare), or from $O(n^2)$ to $O(n)$ for regression methods.}
The different solutions in the decomposition can also be computed fully in parallel, and parallel to the Monte Carlo runs. We report here sequential run times.

From these run times for Case B,  presented in Table \ref{Table: runtimes}, we can see that, using the 2D approximation as a control variate, we gain a computational speed up by a factor 50, as the computation time for the 2D corrections by finite differences is negligible. Using 3D corrections gives us a speed up factor 12; in spite of the greater variance reduction, the increased time for the PDE solutions eats up any benefit over the 2D corrections.

\begin{table}[ht!]
\begin{center}
\caption{Computational times of the individual solvers in seconds. For the finite difference (FD) solvers, we use $m_1 = 60$ grid points and 100 or 500 (in brackets) time steps (see also Appendix \ref{app:numparams}).
The full Monte Carlo (MC) result is obtained with $4 \cdot 10^6$ paths and 100 or 1000 (in brackets) time steps.
For the control variate (CV MC) results, we reduce the number of paths commensurate with the variance reduction. From Figure \ref{fig: var reductions MC}, Case B, the variance reduction factor is seen to be around 50 in the case of 2D and 200 for the 3D corrections. Therefore, we choose $8 \cdot 10^4$ paths in the case of 2D corrections and $2 \cdot 10^4$ for the 3D corrections.}
{\begin{tabular}{{lc}}\toprule
Solver      &  time (s)  \\ \toprule
1D FD       &  1.53 (3.65)  \\
2D FD       & 2.35 (10.67)\\
3D FD       & 344 (601.34)\\ \midrule
Full MC     &  4251 (7738)\\
2D CV MC       & 81.5 (258.4)\\ 
3D CV MC       & 30.1 (107.6)\\ 
\bottomrule
\end{tabular}}
 \label{Table: runtimes}
\label{sample-table}
\end{center}
\end{table}


\subsection{Other base risk factors}
\label{subsec:otherbases}

\begin{figure}[ht!]
\centering
\subfigure[EE for three different three-dimensional risk factor decomposed approximations.]{
\includegraphics[width=0.48\textwidth]{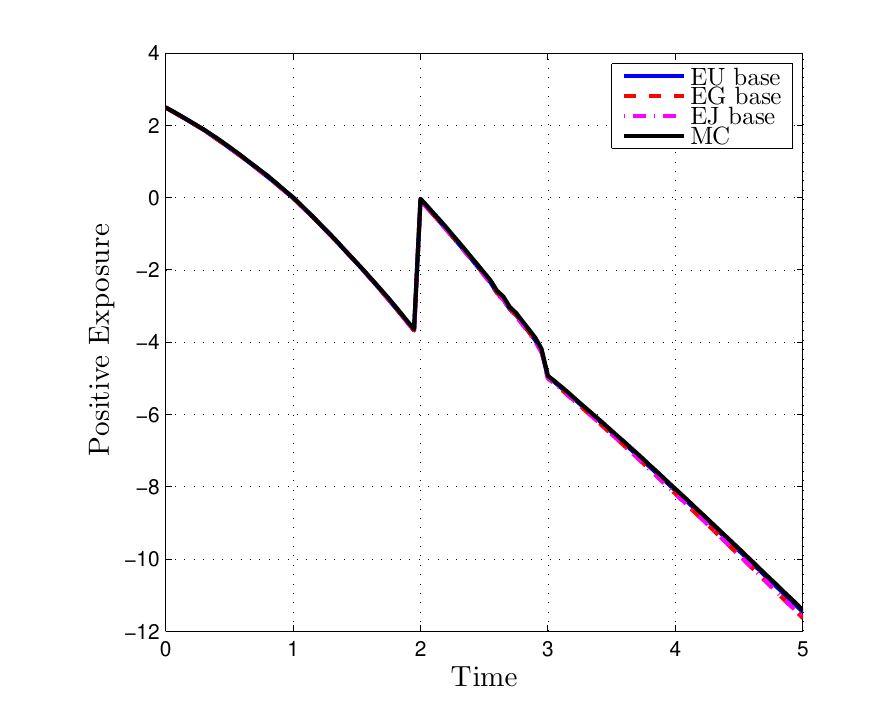}\label{subfig: EE 3D PCAs}
}
\subfigure[EPE for three different three-dimensional risk factor decomposed approximations.]{
\includegraphics[width=0.48\textwidth]{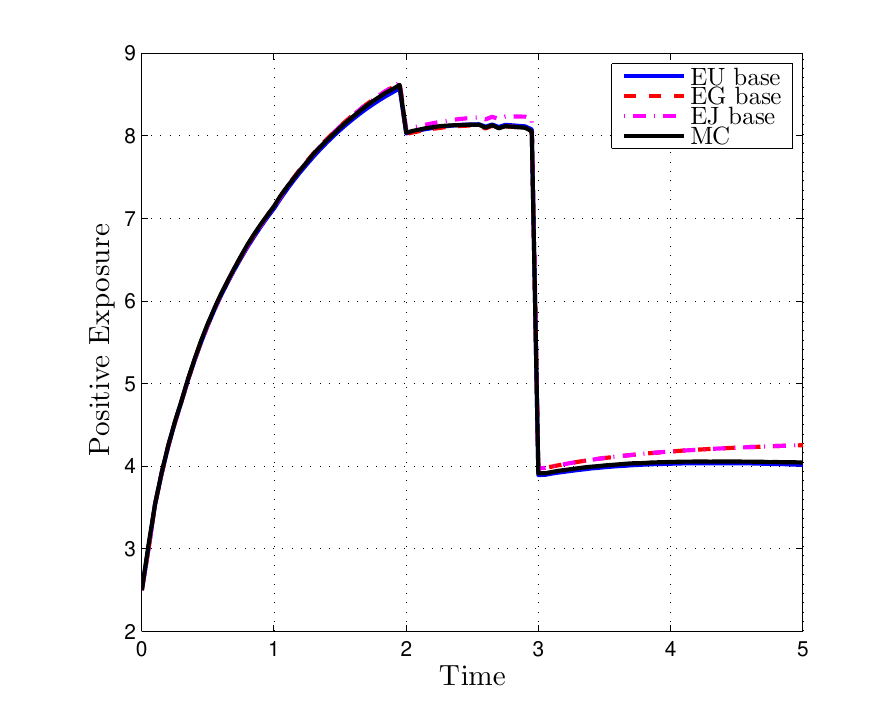}\label{subfig: EPE 3D PCAs}
}
\subfigure[Variance reduction for EE for three different two-dimensional  approximations.]{
\includegraphics[width=0.48\textwidth]{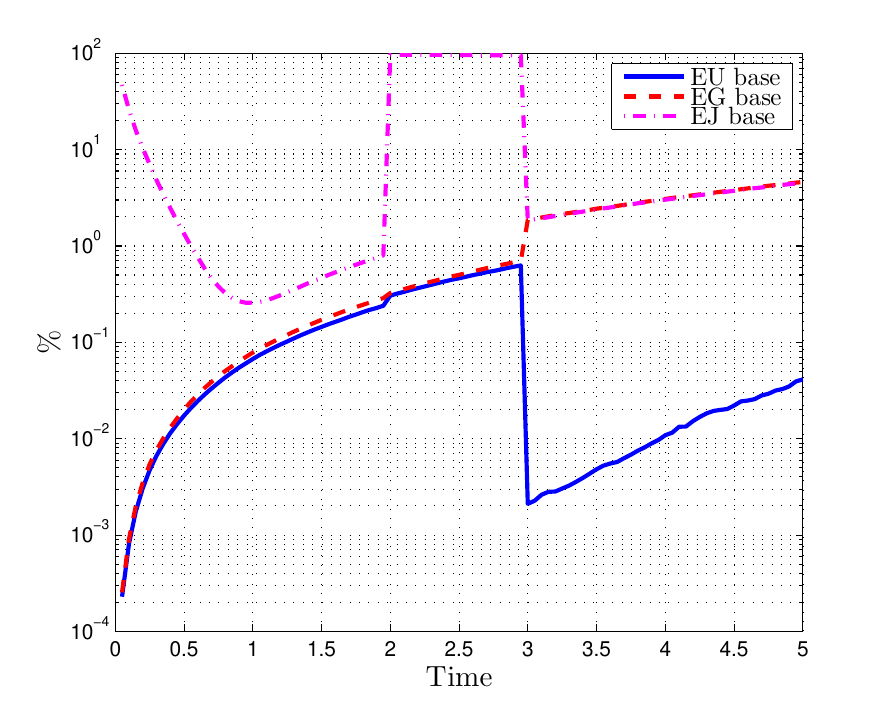}\label{subfig: var reduc PCAs EE 2D}
}
\subfigure[ Variance reduction for EPE for three different two-dimensional  approximations.]{
\includegraphics[width=0.48\textwidth]{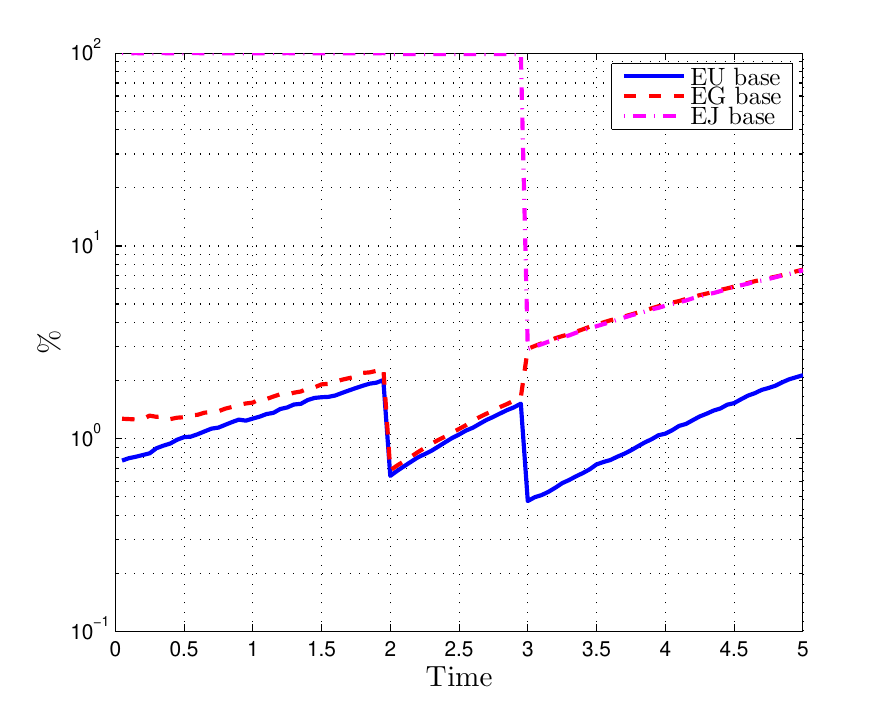}\label{subfig: var reduc PCAs EPE 2D}
}
\centering
\subfigure[ Variance reduction for EE for three different three-dimensional approximations.]{
\includegraphics[width=0.48\textwidth]{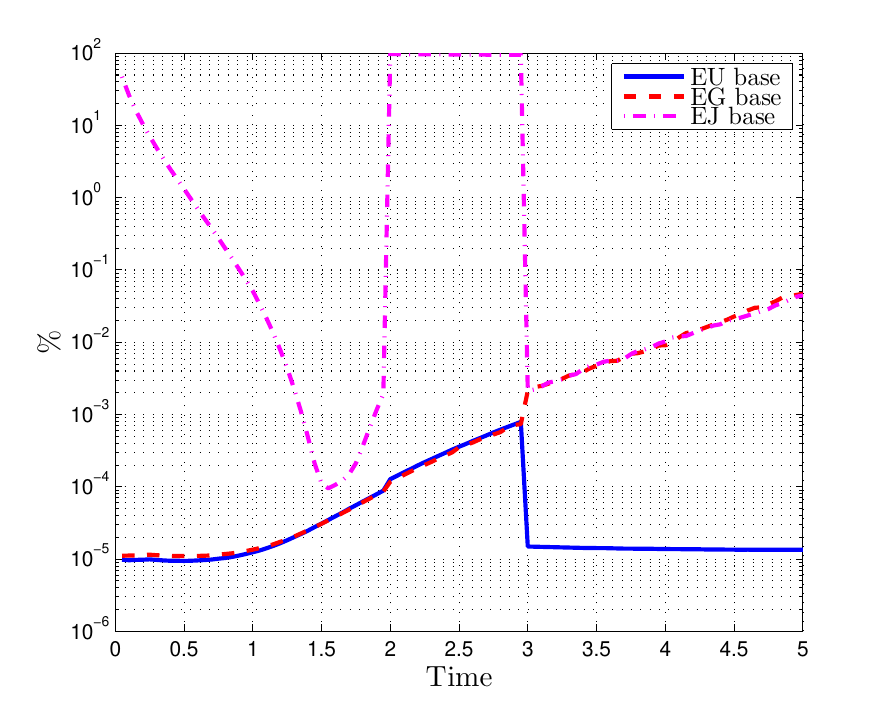}\label{subfig: var reduc PCAs EE 3D}
}
\subfigure[ Variance reduction for EPE for three different three-dimensional  approximations.]{
\includegraphics[width=0.48\textwidth]{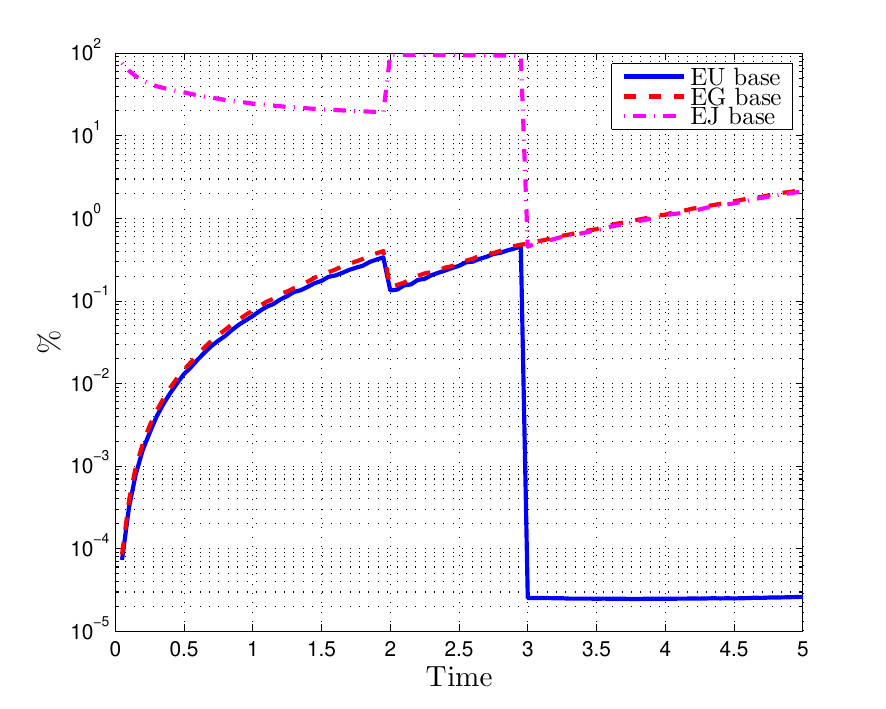}\label{subfig: var reduc PCAs EPE 3D}
}
\caption{
Exposure profiles (EE and EPE) for Case B and variance reduction with different base factors and 2D and 3D corrections. 
For the variance computations $10^6$ paths were used.}
\label{fig: other PCAs}
 \end{figure}

In the previous sections, the EURUSD FX rate was chosen as the base risk factor \emph{a priori}, based on the fact that the derivatives driven by this FX rate have the highest maturity. However, there is no guarantee that this choice as a base is optimal in any sense.
Often, one may have prior knowledge what the main driving factors are. 
Failing that, it would be practically feasible to estimate the variance reduction achieved by different factors by a relatively small number of samples in a trial run, and then do the actual large scale estimation with the best performing base factor.
In Figures \ref{subfig: EE 3D PCAs} and \ref{subfig: EPE 3D PCAs}, the different EE and EPE profiles for Case B are shown when we choose the EURGBP or EURJPY FX rate as a base compared to choosing the EURUSD rate. Table \ref{tab: other PCAs} shows that the error is indeed smallest for the EURUSD rate. 
\begin{table}[ht!]
\begin{center}
\caption{Errors of exposures for Case B for different choices of the base risk factor. The errors are expressed in percentages together with the standard error of the Monte Carlo benchmark with $10^6$ paths. The SE is defined in (\ref{eqn:SE}) as the root sum squared of the standard errors over time relative to the root sum squared of the sampled EE or EPE. \label{tab: other PCAs}}
{\begin{tabular}{{llcc|cc|cc|c}}\toprule
            & &  \multicolumn{2}{c}{$e_{L_2}$}& \multicolumn{2}{c}{$e_{L_\infty}$}& \multicolumn{2}{c}{MD (bp)}& SE \\ \toprule
            & &   2D  & 3D &   2D  & 3D &   2D  & 3D  & \\ \midrule
\multirow{2}{*}{EURUSD} & EE       & 2.02\%& 0.69\% & 1.83\% & 0.59\% & 2.43 & 0.86 &0.23\%    \\
                        & EPE      & 1.97\%& 0.36\% & 2.74\% & 0.51\% & 2.64 & 0.51 &0.15\%\\ \midrule
\multirow{2}{*}{EURGBP} & EE       & 2.32\%& 1.39\% & 2.26\% & 1.60\% & 2.47 & 1.26 &0.23\%    \\
                        & EPE      & 4.01\%& 1.89\% & 4.77\% & 3.05\% & 5.45 & 2.02 & 0.15\%\\ \midrule
\multirow{2}{*}{EURJPY} & EE       & 2.44\%& 1.43\% & 2.26\% & 1.60\% & 2.72 & 1.39 & 0.23\%   \\
                        & EPE      & 7.91\%& 2.03\% & 19.35\% & 3.05\% & 8.95 & 2.46 &0.15\%\\ \bottomrule
\end{tabular}}
\end{center}
\end{table}

In Figures \ref{subfig: var reduc PCAs EE 2D} and \ref{subfig: var reduc PCAs EPE 2D}, we show the resulting variance reductions of the corresponding control variates. Clearly, the EURJPY FX rate as a base does not reduce the variance after $T=2$, as that is when the base risk factor does not affect  any non-terminated derivative in the portfolio. The EURUSD FX rate performs best especially after time $T>3$ when the EURUSD rate affects the only non-terminated derivative in the portfolio. Note that we use a log scale 
for better visualization.


\section{Conclusion}\label{sec: conclusion}


This paper is motivated by the computation of exposure profiles for portfolios depending on a moderate to large number of risk factors. For this problem, the industry standard technique is to employ forward Monte Carlo sampling to compute future scenarios. This scales linearly in both the number of dimensions (risk factors) and products. 
However, valuing the whole book across all scenarios is still a big computational challenge and relatively large standard errors have to be tolerated given the relatively low feasible number of sample paths.

We therefore propose another approach which exploits the accuracy of PDE approximation schemes for low-dimensional estimation problems, through an anchored-ANOVA-style splitting of the high-dimensional problem into a sequence of lower-dimensional ones.
As the problem is stated in the form of nested conditional expectations, we use a combination of forward and backward PDEs to generate exposure profiles for all future times with a single backward and forward sweep.



This paper provides a proof of concept rather than a fully worked out black box algorithm. The detailed analysis of a moderately sized, realistic and fully calibrated test case showed the scale of the benefits achievable by risk factor decomposition coupled with numerical PDE solutions, used as standalone approximation or -- if necessary -- employed as a control variate for a Monte Carlo estimator. Some of the computational savings were dramatic, with a speed-up factor of 50 compared to standard Monte Carlo estimation, by using a sum of two-dimensional estimators as control variate.

Further tests, not reported here, indicate that for longer maturities (10 years) some three-dimensional terms in the decomposition become significant for an accurate standalone approximation as well as for effective variance reduction.

It remains to investigate in practice the scalability with respect to the number of risk factors and number of products in the derivative portfolio. The PDE based method, like standard Monte Carlo sampling, scales linearly in the number of derivatives. As for the number of risk factors, the number of 2D correction terms is linear, and the number of 3D ones quadratic in the dimension.

For larger portfolios, one can consider using approximations of the form $V_{0,2}$ or $V_{0,3}$ instead of $V_{1,2}$, or indeed a combination of the two in the sense that only a subset of $V_u$ with $|u|=2$ and $|u|=3$ are computed and included in the approximation, chosen adaptively by estimation of the individual terms. We find evidence for this in Table \ref{Table: corrections FDs EE}, which shows that only a subset of the correction terms are significant. This also suggests that not all terms have to be computed with the same relative accuracy, and one can follow the principle of \citet{griebel2010dimension} to divide the total computational budget optimally between all correction terms. In this process, it is not necessary to use the same numerical method for all terms. Indeed, we have used a combination of closed-form and numerical solutions in the tests, and the framework is rich enough to use the best available method (e.g., Fourier-based methods for affine models, PDEs for early exercise options, or conceivably even a Monte Carlo method for strongly path-dependent derivatives) for a given sub-problem.
Moreover, for the computation of a specific correction term only a subset of derivatives has to be considered, namely those affected by the risk factors considered in that correction. So it is conceivable that even for a large derivative portfolio each correction term only requires consideration of a small fraction of the derivatives. For instance, if the portfolio contains derivatives on equities, commodities and FX, the risk factor decomposition may provide a way to decompose the exposure computations into smaller sub-computations.

The practical challenge will be to develop a framework which allows this to happen in a generic way, and which is easily adaptable as different derivatives are entered or the modelling framework changes.

%
%

\appendix
\section{Model parameters}\label{app: Parameters}
The full correlation matrix for the process
$\left(X_t^1,\ldots, X_t^7\right) =
\left(
F_t^1, R^{\rm d}, R_t^{{\rm f},1}, F_t^2, R_t^{{\rm f},2}, F_t^3, R_t^{{\rm f},3}
\right)$
from (\ref{XvsFR}),
used in the tests of Section \ref{sec: Results}, after regularisation for positive definiteness \citep{Rebonato1999}:
\begin{eqnarray*}
\begin{pmatrix}
      1 &  -0.3024 &   0.1226  &  0.5815  & -0.0142  &  0.5510  &  0.5351\\
   -0.3024 &   1 &   0.6293  & -0.2577  &  0.6895  & -0.4554  &  0.3188\\
    0.1226 &   0.6293 &   1  &  0.0459  &  0.7453  & -0.3049  &  0.4181\\
    0.5815 &  -0.2577 &   0.0459  &  1  &  0.1230  &  0.5490  & -0.0848\\
   -0.0142 &   0.6895 &   0.7453  &  0.1230  &  1  & -0.3015  &  0.3587\\
    0.5510 &  -0.4554 &  -0.3049  &  0.5490  & -0.3015  &  1  & -0.3260\\
    0.5351 &   0.3188 &   0.4181  & -0.0848  &  0.3587  & -0.3260  &  1
\end{pmatrix}
\end{eqnarray*}
The FX and interest rate SDEs are driven by the parameters in Table \ref{Table: Model paramaters}.
\begin{table}[ht!]
\begin{center}
\caption{Parameters for the Black-Scholes-2-Hull-White (BS2HW) model (\ref{FRprocessesBS}).}
{\begin{tabular}{{lc||lccc}}\toprule
EUR & & & $i=1$ (USD) & $i=2$ (GBP) & $i=3$ (JPY) \\
\toprule
			&			      & $F_0^{i}$ 		&  1.2470 & 0.7926 &  147.53 \\
$R_0^{\rm d}$	 & 1.8157e-04			  & $R_0^{{\rm f},i}$ 	&  -0.0036 &  0.0065 & 0.0011 \\
$\lambda_{\rm d}$		& 0.010				& $\lambda^i_{\rm f}$ & 0.010 &  0.0523 & 0.010 \\
$\eta_{\rm d}$	&		0.0070	  	& $\eta^i_{\rm f}$ 	&  0.0092  & 0.0104 &  0.0057\\ \bottomrule
\end{tabular}}
 \label{Table: Model paramaters}
\end{center}
\end{table}

As discussed in Section \ref{subsec:calibration}, the functions $\Theta_{\it d}(t), \mbox{ and }\Theta^i_{\it f}(t), i=1,2,3$, are calibrated to fit the forward rate curve of the respective markets as seen on 2 December 2014. The ATM volatilities of the FX rates at this date used for $\sigma(t)$ can be found in 
Table \ref{Table: Vaoltilities}.
\begin{table}[ht!]
\begin{center}
\caption{ATM volatilities as seen on 2 December 2014 that are used for bootstrapping the piecewise constant volatility function as explained in Section \ref{subsec:calibration}.}
{\begin{tabular}{{lcccc}}\toprule
 		& EURUSD (\%) & EURGBP (\%)  & EURJPY (\%)\\ \toprule
 $T=1M$ & 8.852  	 &  6.570  	   & 10.247\\
 $T=3M$ & 8.695  	 &  6.635  	   & 10.245\\
 $T=6M$ & 8.580  	 &  7.350  	   & 10.517\\
 $T=1Y$ & 8.605  	 &  7.447  	   & 10.848\\
 $T=2Y$ & 8.717  	 &  7.865  	   & 11.580\\
 $T=3Y$ & 8.952  	 &  8.068  	   & 12.247\\
 $T=5Y$ & 9.635  	 &  8.383  	   & 13.642\\ \bottomrule
\end{tabular}}
 \label{Table: Vaoltilities}
\end{center}
\end{table}

\subsection{Heston model parameters}\label{app: Parameters Heston}

\begin{figure}[h!]
\centering
\subfigure[ EURUSD  ]{
\includegraphics[width=0.48\textwidth]{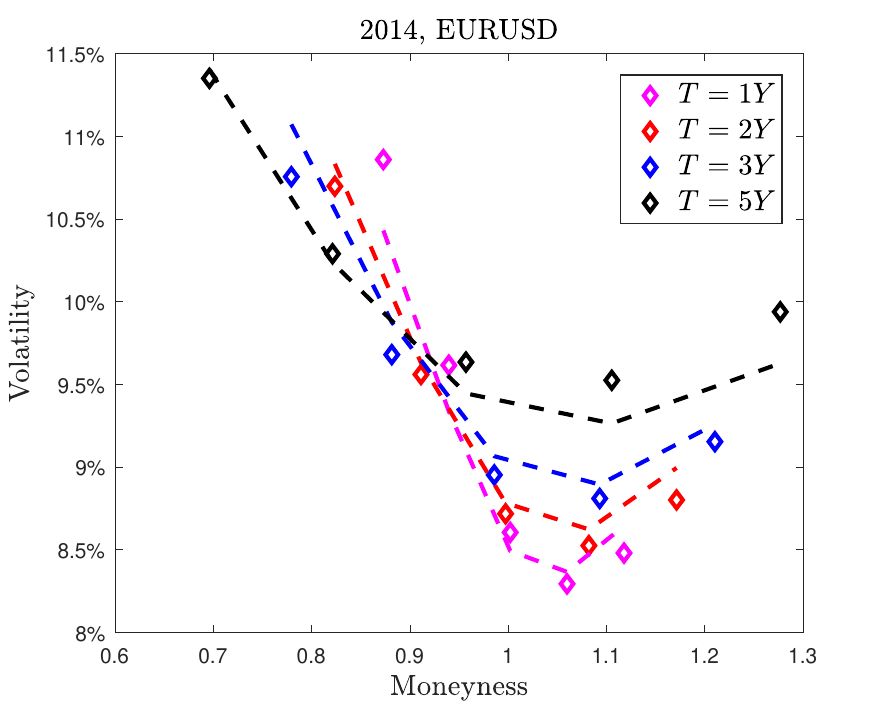}\label{subfig: fit EURUSD}
}
\subfigure[ EURGBP  ]{
\includegraphics[width=0.48\textwidth]{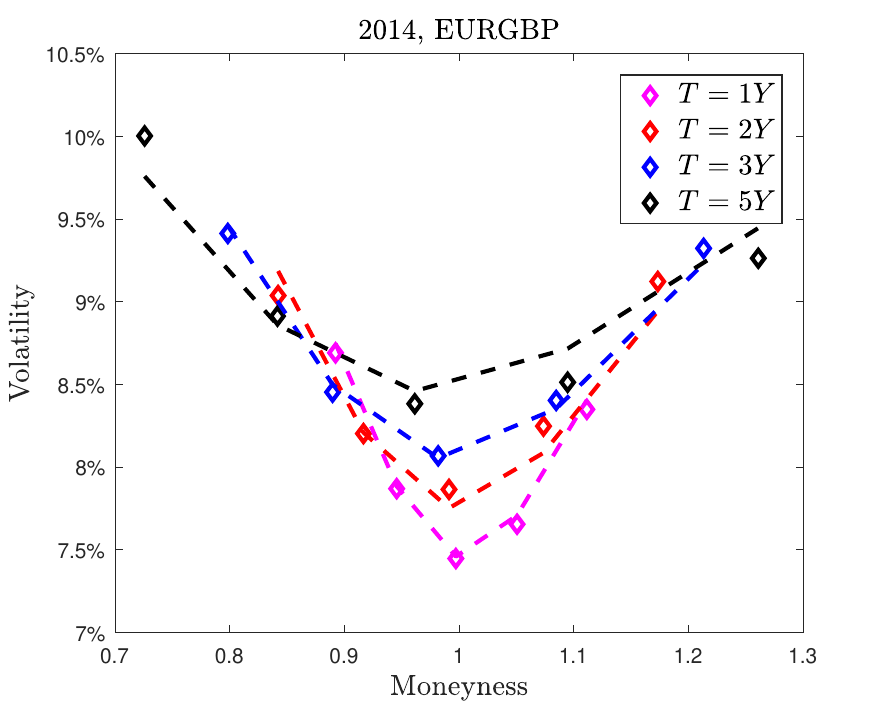}\label{subfig: fit EURGBP}
}
\subfigure[EURJPY   ]{
\includegraphics[width=0.48\textwidth]{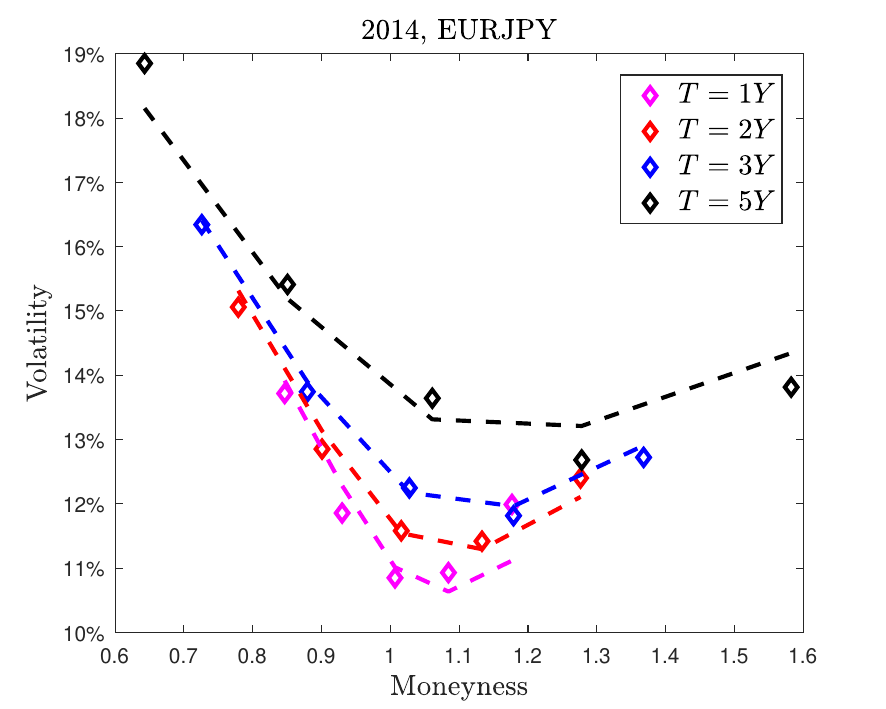}\label{subfig: fit EURJPY}
}
\subfigure[EURJPY   ]{
\includegraphics[width=0.48\textwidth]{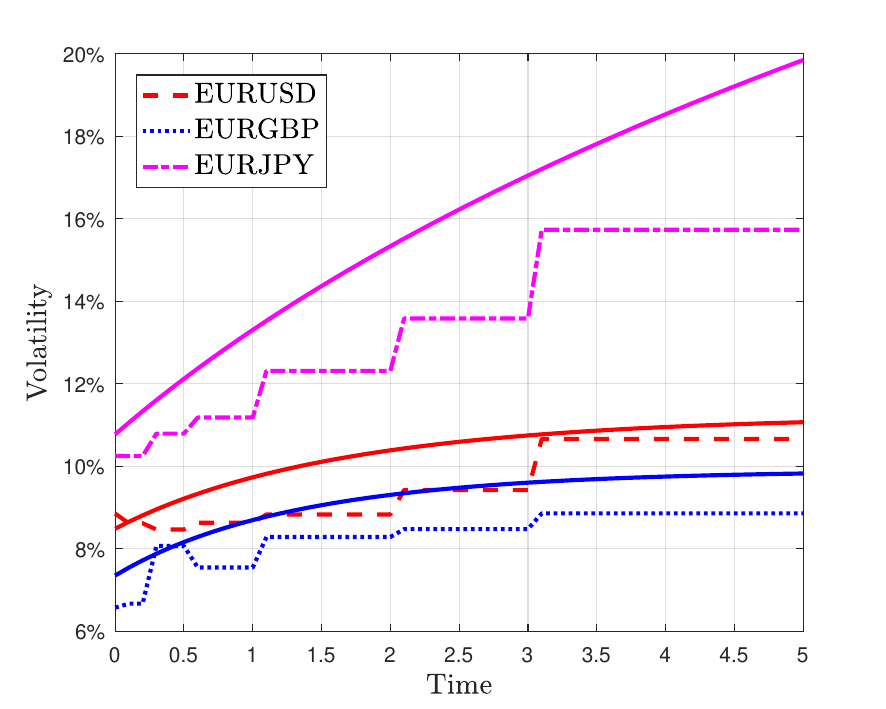}\label{subfig: volatility function}
}
\caption{Implied volatility smiles of FX pairs. The lines are the model fit and the diamonds are the market implied volatilities. The associated errors are presented in Table \ref{Table: calibration details}. In Fig.~\ref{subfig: volatility function}, shown are the piecewise constant volatilities as a function of time, based on bootstrapped ATM volatilities (dashed lines) and the expectations of the model stochastic volatilities 
(solid lines).}
\end{figure}

In Table \ref{Table: calibration details} we give the calibrated Heston parameters together with the mean implied volatility error.
As shown in Figures A\ref{subfig: fit EURUSD}, A\ref{subfig: fit EURGBP} and A\ref{subfig: fit EURJPY}, the skew in the EURJPY market is most pronounced (note the different scales on the y-axis). Due to this skew, the bootstrapped ATM volatility is different from the expectation of the future volatility in (\ref{eq: mean volatility espr}). This mean volatility and the bootstrapped volatility over time are presented in Figure A\ref{subfig: volatility function}.
\begin{table}[ht!]
\begin{center}
\caption{Calibrated Heston parameters to 2014 market data. Including mean implied vol errors quoted in percentages. The models are calibrated to 10\%-, 25\%- and 50\%-$\Delta$ FX put and call options with maturities 1Y, 2Y, 3Y and 5Y.\label{Table: calibration details}}
{\begin{tabular}{{llc|cc}}\toprule
            			& parameters	& 		 &error 	\\ \hline
\multirow{5}{*}{EURUSD} & $\kappa $     & 0.5449 & 1.71\% \\
                        & $v_0 	$		& 0.0072 & \\ 
                        & $\bar{v}$		& 0.0126 & \\ 
                        & $\rho $		& -0.2752& \\ 
                        & $\gamma $		& 0.1560 & \\  \hline
\multirow{5}{*}{EURGBP} & $\kappa $     & 0.6740 & 1.03\% \\
                        & $v_0 	$		& 0.0054 & \\
                        & $\bar{v}$		& 0.0098 & \\  
                        & $\rho $	  	& -0.0762& \\ 
                        & $\gamma $		& 0.1771 & \\  \hline
\multirow{5}{*}{EURJPY} & $\kappa $     &  0.0476&  2.23\% \\
                        & $v_0 	$		& 0.0116 & \\ 
                        & $\bar{v}$		& 0.1428 & \\ 
                        & $\rho $	  	& 0.1890 & \\ 
                        & $\gamma $		& -0.3507& \\  \hline
\end{tabular}}
\end{center}
\end{table}

The full correlation matrix, including stochastic volatility, for the process
$\left(X_t^1,\ldots, X_t^{10}\right) =
\left(
F_t^1,Y_t^1, R^{\rm d}, R_t^{{\rm f},1}, F_t^2,Y_t^2, R_t^{{\rm f},2}, F_t^3,Y_t^3, R_t^{{\rm f},3}
\right)$ used in the tests of Section \ref{subsec:Stochastic Volatility}, after regularisation for positive definiteness \citep{Rebonato1999}:
\begin{eqnarray}
\begin{pmatrix}
    1&   -0.2644&   -0.2910&    0.1142  &  0.5673   &-0.0004 &  -0.0107  &  0.5243 &  -0.0055 &   0.5043\\
   -0.2644&    1&    0.0019&   -0.0018  & -0.0016   &-0.0001 &   0.0009  & -0.0043 &  -0.0014 &  -0.0052\\
   -0.2910&    0.0019&    1&    0.6322  & -0.2558   &-0.0001 &   0.6892  & -0.4535 &  -0.0015 &   0.3284\\
    0.1142&   -0.0018&    0.6322&    1  &  0.0438   & 0.0001 &   0.7473  & -0.3091 &   0.0014 &   0.4171\\
    0.5673&   -0.0016&   -0.2558&    0.0438  &  1   &-0.0740 &   0.1248  &  0.5468 &   0.0013 &  -0.0796\\
   -0.0004&   -0.0001&   -0.0001&    0.0001  & -0.0740   & 1 &  -0.0001  &  0.0003 &   0.0001 &   0.0003\\
   -0.0107&    0.0009&    0.6892&    0.7473  &  0.1248   &-0.0001 &   1  & -0.3001 &  -0.0007 &   0.3659\\
    0.5243&   -0.0043&   -0.4535&   -0.3091  &  0.5468   & 0.0003 &  -0.3001  &  1 &  -0.3452 &  -0.3103\\
   -0.0055&   -0.0014&   -0.0015&    0.0014  &  0.0013   & 0.0001 &  -0.0007  & -0.3452 &   1 &   0.0041\\
    0.5043&   -0.0052&    0.3284&    0.4171  & -0.0796   & 0.0003 &   0.3659  & -0.3103 &   0.0041 &   1  
\end{pmatrix}\label{eq: Heston correlations}
\end{eqnarray}
Note, that due to the regularisation the correlation matrix is altered, and correlations are slightly different from the Black-Scholes case.

\section{Numerical approximation of the 
Kolmogorov PDEs}
\label{sec:findiff}

Here, we outline the finite difference method used for the PDE solutions, as it is somewhat non-standard through the use of forward and backward equations. In particular, we extend the adjoint method from \citet{Itkin2014} from two to three dimensions and use it for the forward component.
For a more general introduction to finite difference methods for pricing financial derivatives, see for example \citet{TavellaRandall2000} and, specifically for interest rate derivatives \citet{AndersenPiterbarg2010}. 

As usual, a grid is defined with one dimension per risk factor, and the partial derivatives in the Kolmogorov forward (\ref{eq: Forward Kolmogorov PDE})  or backward PDE (\ref{eq: Backward Kolmogorov PDE}) are approximated by finite differences on this grid.

\subsection{Backward equations}

We use a combination of second order central differences in the centre of the spatial domain, and upwinding for stabilisation of large drifts in outer parts of the domain (see Appendix \ref{app:numparams}), to obtain a system of ODEs in the time variable~\citep{Hout2009}. Let $U(\tau)$ be the solution to this semi-discrete Kolmogorov backward equation at time $\tau$, where $\tau = T-t$, then we have the following initial value problem
\begin{eqnarray}
\label{semi-discrU}
\frac{\textrm{d} U}{\textrm{d} \tau} = F(\tau) U,  \quad \tau \geq 0, \qquad U(0) = U_0, 
\end{eqnarray}
with given matrix 
$F$, derived from the PDE, and initial vector $U_0$ given by the payoff at the mesh points. 
The boundary conditions are also derived from the payoff function.

We apply an Alternating Direction Implicit (ADI) splitting method.
Let $F^n=F(\tau_n)$ be the discretisation matrix, as per (\ref{semi-discrU}), at time step $\tau_n=n \Delta \tau$, $n\in \mathbb{N}$, $\Delta \tau>0$ a uniform step size,  then this matrix is first decomposed into
\begin{eqnarray*}
F^n = F_0^n +F_1^n+F_2^n+F_3^n, 
\end{eqnarray*}
where the individual $F^n_i, 1\leq i\leq d$ contains the contribution to $F$ stemming from the first and second order derivatives in the $i$th dimension. 
Following \citet{intHout2010}, we define one matrix  $F_0$ which accounts for the mixed derivative terms, and treat that fully explicitly.
Here we show the scheme for a three-dimensional problem, and the two-dimensional case is obtained as special case by setting $F_3=0$.
The Hundsdorfer-Verwer (HV) scheme \citep{HundsdorferVerwer2003},
\begin{eqnarray*}
Y_0 &= U_{n-1}+\Delta\tau F^{n-1}U_{n-1},\qquad\qquad\qquad\qquad\quad\\
(I-\theta \Delta \tau F^n_j)Y_j&=Y_{j-1}-\theta\Delta\tau F_j^{n-1}U_{n-1}, \qquad \qquad j=1,2,3,\\
\tilde{Y}_0&=Y_0+\frac{1}{2}\Delta\tau\left[F^nY_3-F^{n-1}U_{n-1})\right],\qquad\quad\quad\\
(I-\theta \Delta \tau F^n_j)\tilde{Y}_j&=Y_{j-1}-\theta\Delta\tau F_j^{n}Y_j, \qquad\quad\qquad j=1,2,3,\\
U_n&=\tilde{Y}_3,\qquad\quad\quad\qquad\quad\quad\qquad\quad\quad\qquad\quad\quad\\
\end{eqnarray*}
defines a second order consistent ADI splitting for all $\theta$, and can be shown to be \emph{von Neumann} stable for $\theta\in\left[ \frac{1}{2} + \frac{1}{6}\sqrt{3},1\right]$, see~\citet{Haentjens2011}. We use $\theta  = 0.8  > \frac{1}{2} + \frac{1}{6}\sqrt{3} \approx 0.789$ in the computations. Accuracy and stability do not appear to be very sensitive to the choice of $\theta$ within the above range.

\subsection{Forward equations}

We use the adjoint relation between the Kolmogorov forward and the backward PDE. It can easily be seen that (for sufficiently smooth coefficients)
\begin{eqnarray}
\label{fwdODE}
\frac{\textrm{d} P}{\textrm{d} t} = F^{\T}(t) P, \quad t\geq 0, \qquad P(0)=P_0,
\end{eqnarray}
with $F$ from above, is a consistent scheme for the Kolmogorov forward equation. The initial datum for the discrete density function $P$ is given by an approximation $P_0$ to the Dirac delta. 
We choose a mesh such that a mesh point coincides with the location of the Dirac delta. Then $P_0$ is set to zero for all other mesh points and to a large value at this particular point, chosen such that a numerical quadrature rule applied to $P_0$ gives 1.
At the spatial boundaries, we apply zero Dirichlet conditions.

Hence, following \citet{Andreasen2010} and \citet{Capriotti2015}, we can first set up matrices that approximate the partial derivatives in the Kolmogorov Backward PDE, and use the transpose of these matrices for the Kolmogorov forward PDE.

Now we could apply HV splitting to (\ref{fwdODE}) and obtain a second order (in time) consistent approximation to the forward PDE. Instead, we follow \citet{Itkin2014} to calculate the exact adjoint of the HV scheme for the backward equation, which results in a different scheme, i.e., the transposition and approximate factorisation do not commute.
The forward scheme, adapted to the three-dimensional case from the two-dimensional case analysed in \citet{Itkin2014}, then is
\begin{eqnarray*}
(I-\theta\Delta t F_3^n)^{\T} Y_0=&P_{n-1},\hspace{10cm}\\
(I-\theta\Delta t F_2^n)^{\T} Y_1=&Y_0,\hspace{10cm}\\
(I-\theta\Delta t F_1^n)^{\T} Y_2=&Y_1,\hspace{10cm}\\
\tilde{Y}_0=& P_{n-1} + \Delta t \left( \left( \frac{1}{2}(F^n)^{\T} - \theta (F_1^n)^{\T} \right) Y_2  - \theta  (F_2^n)^{\T}  Y_1 - \theta  (F_3^n)^{\T}  Y_0\right)\hspace{0.5cm}\\
(I-\theta\Delta t F_3^n)^{\T}\tilde{Y}_1=&\tilde{Y}_0,\hspace{10cm}\\
(I-\theta\Delta t F_2^n)^{\T}\tilde{Y}_2=&\tilde{Y}_1,\hspace{10cm}\\
(I-\theta\Delta t F_1^n)^{\T}\tilde{Y}_3=&\tilde{Y}_2,\hspace{10cm}\\
P_{n}=&\tilde{Y}_3 + \Delta t \left( ( (F^{n-1})^{\T} - \theta  (F_1^{n-1})^{\T})  (\tilde{Y}_3 - Y_2)- \theta (F_2^{n-1})^{\T}(\tilde{Y}_2 - Y_1)\right. \\
&\left.-\theta (F_3^{n-1})^{\T} (\tilde{Y}_1 - Y_0)+ \frac{1}{2}(F^{n-1})^{\T} \tilde{Y}_3\right).
\end{eqnarray*}
By using this scheme, we have the adjoint relation
\begin{eqnarray*}
P_0^{\T} U_n = P_n^{\T} U_0,
\end{eqnarray*}
i.e., using the backward equation to compute the conditional expectation function and then evaluating it at the location of the Dirac measure (left-hand side), or computing the density and then integrating over the ``payoff'' function, gives exactly the same result.
Apart from the mathematical aestheticism of using the exact adjoint, this scheme has the following stability advantages. As discussed in \citet{Wyns2016},
splitting schemes do not give stable, and hence convergent, solutions for Dirac initial data. A direct splitting of (\ref{fwdODE}) is prone to wildly oscillating densities and may give meaningless expectations. By using the above scheme instead, we can be sure that even if we have no guarantee for the stability of the solution, the derived quantity of interest, e.g., EPE, is the same as for the backward equation with regular (for EPE, Lipschitz and piecewise smooth) data.\footnote{But see Remark \ref{rem:fwdbwd} why we are not using the the backward equation directly to compute exposures.}
The adjoint equation also ensures conservation of discrete probability in the interior of the mesh. The loss of mass at the boundaries is minimised by choosing the domain large enough.

With the grid of discrete transition probabilities, obtained by the numerical solution of the Kolmogorov forward PDE, and a grid of portfolio values from the backward PDE approximations, the exposures can be extracted by numerical quadrature applied to (\ref{integral}).

\subsection{Grid construction and numerical parameters}
\label{app:numparams}


The grids used for the approximation 
are non-uniform as in \citet{Haentjens2013}, where a $\sinh$ transformation is used to place a high density of grid points  around the spot values. Moreover, the grid is shifted to include the initial spot value where the non-smooth Dirac delta function for the forward equation is located.

In Table \ref{Table: FD paramaters} the computational domain  in $F_t$, $R^\text{d}_t$ and $R^{\text{f}}_t$ is chosen. The parameter $\xi$ controls the fraction of points that lie close to initial spot values (see \citet{Haentjens2013}). 
\begin{table}[ht!]
\begin{center}
\caption{Finite difference grid parameters.}
{\begin{tabular}{{lccc}}\toprule
                          &min &max & $\xi$ \\\toprule
FX rate ($F_t$)        &0 &$8F_0$ & 20\\
Domestic IR ($R^\text{d}_t$)  &-0.5 &0.8 & 100\\
Foreign IR ($R^{\text{f}}_t$)   &-0.5 &0.8 & 100\\ \bottomrule
\end{tabular}}
 \label{Table: FD paramaters}
\end{center}
\end{table}

Moreover, we use upwind differences for large absolute interest rates, i.e., $R\in [-0.5,-0.1] \cup [0.2,0.8]$, otherwise central differences.

The number of grid points $m_1$ $m_2$, $m_3$ in the three directions is chosen as $m_1 = 2 m_2 = 2 m_3$; 
see also Appendix \ref{app: FD Errors}.

For the $F$ grid, an interval $[F_{\text{\scriptsize left}},F_{\text{\scriptsize right}}]\subseteq [F_{ \text{\scriptsize min}},F_{\text{\scriptsize max}}]$ is defined wherein the grid is uniform and dense, similar to~\citet{Haentjens2011}, while outside the $\sinh$ transformation is used.
We set this interval as $[F_{\text{\scriptsize left}},F_{\text{\scriptsize right}}]=[0.95F_0, 1.02F_0]$.

\section{Regression-based Monte Carlo algorithm}
\label{sec:regression}

We use constant, linear and bi-linear basis functions 
\[
\{\psi_j : j=1,\ldots, 6\} = \{F, R^{\rm d}, R^{\rm f}, F R^{\rm d}, F R^{\rm f}, R^{\rm d} R^{\rm f}\},
\]
to account for the correlation between risk factors.
The algorithm determines option values $Y_i(t)= Y(t;{\omega}_k)$ along the sample path ${\omega}_i$, and can be summarised by the following steps:
\begin{enumerate}
\setcounter{enumi}{-1}
\item Generate all the paths $(F_t({\omega}_i), R_t^{\rm d}({\omega}_i), R_t^{\rm f}({\omega}_i))$ at all the time points $t=0,\Delta t, \ldots$.
\item Start from maturity and set $t \leftarrow  T$ , $Y_i(T)  \leftarrow \phi({F}_T({\Nom}_i))$.
\item Set $Y_i(t) \leftarrow  \exp(- \Delta t R^{\rm d}_t({\omega}_i)) Y_i(t)$. 
\item  
Solve the linear regression problem
\begin{eqnarray*}
\mathbb{E}\left[\Big( \mathbb{E}\Big[Y(t) \Big \vert X_{t-\Delta t} \right] - \sum_{j=1}^6 \beta_j \psi_j(X_{t-\Delta t}) \Big)^2 \Big]
\quad \longrightarrow \quad \min_\beta
\end{eqnarray*}
by approximating the expectations on the paths ${\omega}_i$ to find a minimiser $\hat{\beta}$.
\item
Set $t \leftarrow T-\Delta t$ and
\begin{eqnarray*}
Y_i(t)  \leftarrow& \;\, \hat{\beta}_0 + \hat{\beta}_1 F_t({\omega}_i) + \hat{\beta}_2 R_t^{\rm d}({\omega}_i)+ \hat{\beta}_3 R_t^{\rm f}({\omega}_i)\hspace{3cm}\\
&+ \hat{\beta}_4 F_t({\omega}_i) R_t^{\rm d}({\omega}_i) + \hat{\beta}_5 F_t({\omega}_i)R_t^{\rm f}({\omega}_i)+
\hat{\beta}_6 R_t^{\rm d}({\omega}_i)R_t^{\rm f}({\omega}_i)
\end{eqnarray*}
\item Repeat backwards from 2.\ for all time steps.
\end{enumerate}

\section{Finite difference errors and individual terms for Case B}\label{app: FD Errors}

Here we show the individual FD errors for all terms involved for Case B, compared to a Monte Carlo estimate,
for both EE (Table \ref{Table: error FDs EE}) and EPE (Table \ref{Table: error FDs EPE}).

\begin{remark}
\label{nomenclature}
In the following tables, for instance, BSHW EU-RU refers to $V_{\{1,5\}}(F^{1},R^{{\rm f},1})$ from Section \ref{sec: Results}, i.e.,
a Black-Scholes-Hull-White model where the EURUSD rate $F^1$ follows a Black-Scholes-type model but with Hull-White foreign short rate $R^{{\rm f},1}$ and deterministic domestic rate,
and similarly for the other terms.
\end{remark}


\begin{table}[ht!]
\begin{center}
\caption{Finite difference errors of exposures (EE) of a portfolio with three CCYS. The nomenclature of the terms is explained in Remark \ref{nomenclature}. The errors are measured in $e_{L_2}$ (see equation (\ref{eq: eL2})) in percentages for different number of mesh points. The number of time steps is fixed at 500. 
Within brackets (for $m_1=40$) the error for a FD solution with 1000 time steps.
 The Monte Carlo benchmark is computed with $4\cdot10^6$ paths and 1000 time steps.}
{\begin{tabular}{{lccccc}}\toprule
            	& MC SE& $m_1 = 40$&$m_1 = 60$&$m_1 = 80$ & $m_1 = 100$     \\ \toprule
BS EU           & 0.15   & 0.22 (0.18)    & 0.064 & 0.12& 0.077  \\
BSBS EU-EG      & 0.22   & 0.23 (0.20)    & 0.072  & 0.13 & 0.087 \\
BSBS EU-EJ      & 0.22   & 0.22 (0.19)    & 0.050  & 0.10 & 0.057 \\
BSHW EU-RE      & 0.21   & 0.24 (0.15)    & 0.15   & 0.20  & 0.17 \\
BSHW EU-RU      & 0.22   & 0.22 (0.20)    & 0.23   & 0.18  & 0.21 \\
BSHW EU-RG      & 0.22   & 0.22 (0.18)    & 0.058  & 0.11  & 0.071 \\
BSHW EU-RJ      & 0.22   & 0.21 (0.17)    & 0.061   & 0.11  & 0.075 \\ \midrule
BSBSBS EU-EG-EJ & 0.21   & 0.24 (0.20)    & 0.088  & 0.14  & 0.10 \\
BSHWHW EU-RE-RU & 0.21   & 0.16 (0.14)    & 0.14   & 0.098  & 0.11 \\
BSBSHW EU-EG-RE & 0.21   & 0.23 (0.15)    & 0.16   & 0.21  & 0.18 \\
BSBSHW EU-EJ-RE & 0.21   & 0.22 (0.15)    & 0.13   & 0.18  & 0.15 \\
BSBSHW EU-EG-RU & 0.22   & 0.24 (0.23)    & 0.21   & 0.17  & 0.19 \\
BSBSHW EU-EJ-RU & 0.22   & 0.22 (0.21)    & 0.20   & 0.16  & 0.18 \\
BSBSHW EU-EG-RG & 0.22   & 0.25 (0.22)    & 0.088   & 0.14  & 0.096 \\
BSBSHW EU-EJ-RJ & 0.22   & 0.23 (0.19)    & 0.063   & 0.12  & 0.074	\\ \bottomrule
\end{tabular}}
 \label{Table: error FDs EE}
\end{center}
\end{table}

\begin{table}[ht!]
\begin{center}
\caption{Finite difference errors of positive exposures (EPE) of a portfolio with three CCYS. The nomenclature of the terms is explained in Remark \ref{nomenclature}. The errors are measured in $e_{L_2}$ (see equation (\ref{eq: eL2})) in percentages for different number of mesh points. The number of time steps is fixed at 500. Within brackets (for $m_1=40$) the error for a FD solution with 1000 time steps. The Monte Carlo benchmark is computed with $4\cdot10^6$ paths and 1000 time steps.  }
{\begin{tabular}{{lccccc}}\toprule
            	& MC SE& $m_1 = 40$&$m_1 = 60$&$m_1 = 80$ &$m_1 = 100$    \\ \toprule
BS EU           &  0.073 & 0.30 (0.34)    & 0.099   &  0.042   & 0.030   \\
BSBS EU-EG      &  0.070 & 0.23 (0.25)    & 0.11   &  0.078  & 0.071  \\
BSBS EU-EJ      &  0.084 & 0.24 (0.27)    & 0.070   &  0.038   & 0.048  \\
BSHW EU-RE      &  0.092 & 0.31 (0.37)    & 0.11   &  0.050   & 0.041  \\
BSHW EU-RU      &  0.094 & 0.45 (0.42)    & 0.24   &  0.17   & 0.14  \\
BSHW EU-RG      &  0.093 & 0.29 (0.32)    & 0.090   &  0.040   & 0.043 \\
BSHW EU-RJ      &  0.093 & 0.30 (0.34)    & 0.10   &  0.054   & 0.048 \\ \midrule
BSBSBS EU-EG-EJ &  0.064 & 0.21 (0.23)    & 0.090   &  0.056   & 0.045\\
BSHWHW EU-RE-RU &  0.091 & 0.53 (0.51)    & 0.31   & 0.24    & 0.21\\
BSBSHW EU-EG-RE &  0.069 & 0.22 (0.25)    & 0.12   &  0.10   & 0.10 \\
BSBSHW EU-EJ-RE &  0.083 & 0.25 (0.29)    & 0.080   &  0.054   & 0.062  \\
BSBSHW EU-EG-RU &  0.071 & 0.33 (0.30)    & 0.18   & 0.13    & 0.11 \\
BSBSHW EU-EJ-RU &  0.085 & 0.44 (0.41)    & 0.26   & 0.20    & 0.17 \\
BSBSHW EU-EG-RG &  0.070 & 0.26 (0.28)    & 0.12   &  0.076   & 0.058 \\
BSBSHW EU-EJ-RJ &  0.084 & 0.29 (0.32)    & 0.11   &  0.051   & 0.030 	\\ \bottomrule
\end{tabular}}
 \label{Table: error FDs EPE}
\end{center}
\end{table}

The results in Table \ref{Table: error FDs EE} (especially) and Table \ref{Table: error FDs EPE} show that the FD errors are already in the same order of magnitude as the MC standard error for a larger number of samples (i.e., significantly more samples than are used in practice). We therefore use $m_1 = 60$ mesh points and $500$ time steps in most of the computations (unless otherwise stated).

In Table \ref{Table: error FD and MCs}, we report the accuracy of the complete approximations for increasing number of mesh points $m_1$.
For reference, we also compute the approximation with a standard Monte Carlo estimator.

\begin{remark}
\label{two MC estimators}
In the first column of Table \ref{Table: error FD and MCs}, we report the results for an estimator where the same Brownian paths are used for the estimation of all $V_w$ in (\ref{surplus}) for a given $\Delta V_u$, and in brackets the results if the same paths are also used across all $u$.
It does not seem generally clear which of the estimators has the smaller variance. Using independent paths for different $u$ results in a summation of the variances of $\Delta V_u$. Using the same paths for different $u$ is expected to increase the variance if the terms are predominantly  positively correlated, and decrease the variance if they are predominantly negatively correlated.
\end{remark}

\begin{table}
\begin{center}
\caption{Errors of decomposed approximations to exposures of a portfolio with three CCYS (see Table \ref{Table: error table}, Case B), computed with different numbers of mesh points. The errors are measured in $e_{L_2}$ (see equation (\ref{eq: eL2})) in percentages for different number of grid points in space. The number of time steps is fixed at 500. The Monte Carlo estimator for these approximation errors is computed with $4\cdot10^6$ paths and 1000 time steps. Within brackets for MC the error for a MC simulation with identical paths for all estimators, as explained in Remark \ref{two MC estimators}. Within brackets for $m_1=40$ the error for a FD computation with 1000 timesteps.  }
{\begin{tabular}{{llcccccc}}\toprule
            	      & 			  & MC 	    &$m_1 = 40$ & $m_1 = 60$ & $m_1 = 80$ & $m_1 = 100$ \\ \toprule
\multirow{2}{*}{EE}& 2D corrections   & 1.27 (1.27)  & 1.35 (1.31) 	& 1.41     & 1.36     & 1.39 \\
				   & 3D corrections   & 0.19 (0.24)  & 0.25 (0.27)    & 0.20     & 0.18     & 0.20\\ \midrule			
\multirow{2}{*}{EPE}&2D corrections   & 2.39 (2.38)  & 2.20 (2.21)    & 2.29     & 2.32	  & 2.36\\	
					&3D corrections   & 0.41 (0.24)  & 0.31 (0.29)	& 0.26     & 0.26     & 0.26\\ \bottomrule
\end{tabular}}
 \label{Table: error FD and MCs}
\end{center}
\end{table}

\label{sec:corrections}

\begin{table}
\begin{center}
\caption{Contribution of individual corrections for EE and EPE of a portfolio with three CCYS. 
The nomenclature of the terms is explained in Remark \ref{nomenclature}.
The differences are measured in $e_{L_2}$ (see equation (\ref{eq: eL2})) in percentages for $m_1 = 60$ grid points in space. }
\begin{minipage}{60mm}
\begin{tabular}{{lcc}}\toprule
{ $\quad$ 2D terms  }          	&EE       & EPE \\\toprule
BSBS EU-EG        & 14.58 &  40.35     \\
BSBS EU-EJ        & 8.74  &  18.40     \\
BSHW EU-RE        & 1.40  &   2.73    \\
BSHW EU-RU        & 0.52  &   0.51   \\
BSHW EU-RG        & 0.0003& 0.0022   \\
BSHW EU-RJ        & 0.000 & 0.0093     \\ 
&&\\
&& \\\bottomrule
\end{tabular}
\end{minipage}
\hspace{1 cm}
\begin{minipage}{60mm}
\begin{tabular}{{lcc}}\toprule
{ $\qquad$ 3D terms   }         	&EE       & EPE \\\toprule
BSBSBS EU-EG-EJ   & 0.000 &   0.50  \\
BSHWHW EU-RE-RU   & 1.19  &   2.37  \\
BSBSHW EU-EG-RE   & 0.31  &   1.28  \\
BSBSHW EU-EJ-RE   & 0.18  &   0.72    \\
BSBSHW EU-EG-RU   & 0.000 &  0.083  \\
BSBSHW EU-EJ-RU   & 0.000 &   0.58  \\
BSBSHW EU-EG-RG   & 0.046 &  0.066    \\
BSBSHW EU-EJ-RJ   & 0.12  &   0.30 \\\bottomrule
\end{tabular}
 \label{Table: corrections FDs EE}
 \end{minipage}
\end{center}
\end{table}


\section*{Acknowledgement}
The authors would like to thank Dr.~Sumit Sourabh for helpful advice and Dr.~Shashi Jain for providing the data. Furthermore, financial support by the Dutch Technology Foundation STW (project 12214) is gratefully acknowledged.

%
%



\bibliographystyle{rQUF}
\bibliography{DimReduc_revision}

\end{document}